\documentclass[showpacs,aps,twocolumn]{revtex4}
\usepackage{epsfig}
\usepackage{graphicx}
\usepackage{amsmath,amssymb,amsfonts}
\usepackage{array}
\usepackage{url}
\usepackage{hyperref}
\usepackage{subfig}
\usepackage{multirow}
\usepackage{float}
\usepackage{lineno}
\usepackage{xspace}
\usepackage[usenames,dvipsnames]{color}

\newcommand{\bef}{\begin{figure}}
\newcommand{\eef}{\end{figure}}
\newcommand{\bc}{\begin{center}}
\newcommand{\ec}{\end{center}}
\newcommand{\nn}{\nonumber}
\newcommand{\be}{\begin{equation}}
\newcommand{\ee}{\end{equation}}
\newcommand{\bea}{\begin{eqnarray}}
\newcommand{\eea}{\end{eqnarray}}

\begin{document}
\title{Indication of a Differential Freeze-out in Proton-Proton and Heavy-Ion Collisions at RHIC and LHC energies}
\author{Dhananjaya Thakur}
\author{Sushanta Tripathy}
\author{Prakhar Garg\footnote{Current Affiliation: Department of Physics and Astronomy, Stony Brook University, SUNY, Stony Brook, New York 11794-3800, USA}}
\author{Raghunath Sahoo\footnote{Corresponding author: $Raghunath.Sahoo@cern.ch$}}
\affiliation{Discipline of Physics, School of Basic Sciences, Indian Institute of Technology Indore, Khandwa Road, Simrol, Madhya Pradesh- 453552, India.}
\author{Jean Cleymans}
\affiliation{UCT-CERN Research Centre and Department of Physics, University of Cape Town, Rondebosch 7701, South Africa}

\begin{abstract}
  The experimental data from the RHIC and LHC experiments of invariant  $p_{\rm T}$ spectra for most peripheral A+A and $p+p$ collisions are analyzed with Tsallis distributions in different approaches. The information about the freeze-out surface in terms of freeze-out volume, temperature, chemical potential and radial flow velocity for $\pi^{+}$,  $K^{+}$, p and their anti-particles are obtained. Further, these parameters are studied as a function of the mass of the particles. A mass-dependent differential freeze-out is observed which does not seem to distinguish between particles and their antiparticles. Furthermore, a mass-hierarchy in the radial flow is observed, meaning heavier particles suffer lower radial flow. Tsallis distribution function at finite chemical potential is used to study the mass dependence of chemical potential. The peripheral heavy-ion and proton-proton collisions at the same energies seem to be equivalent in terms of the extracted thermodynamic parameters.
\end{abstract}
\date{\today}
\maketitle

\section{Introduction}
\label{intro}
  High-energy heavy ion collisions provide an unique opportunity to study the nuclear matter under extreme conditions i.e. at high temperature and/or density. Due to high multiplicities produced in A+A  and $p+p$ collisions, the statistical models are more suitable to describe the particle production mechanism. Such a statistical description of transverse momentum $(p_{T})$ of final state particles produced in high-energy collisions has been proposed to follow a thermalized Boltzmann type of distribution as given by  \cite{Hagedorn:1965st}

         \begin{eqnarray}
         \label{eq1}
         E\frac{d^3\sigma}{d^3p}& \simeq C \exp\left(-\frac{p_T}{T_{exp}}\right).
         \end{eqnarray}
         
  To account for the high-$p_{\rm T}$ tail, a power-law in $p_{\rm T}$ has been proposed \cite{CM,UA1}, which empirically accounts for the possible QCD contributions. Hagedorn proposed a combination of both the aspects, which describes the experimental data over a wide  $p_{\rm T}$ range  \cite{Hagedorn:1983wk} and is given by 
  
\begin{eqnarray}
  E\frac{d^3\sigma}{d^3p}& = &C\left( 1 + \frac{p_T}{p_0}\right)^{-n}
\nonumber\\
 & \longrightarrow&
  \left\{
 \begin{array}{l}
  \exp\left(-\frac{n p_T}{p_0}\right)\quad \, \, \, {\rm for}\ p_T \to 0, \smallskip\\
  \left(\frac{p_0}{p_T}\right)^{n}\qquad \qquad{\rm for}\ p_T \to \infty,
 \end{array}
 \right .
 \label{eq2}
\end{eqnarray}

where $C$, $p_0$, and $n$ are fitting parameters. This becomes a purely exponential function for small $p_{\rm T}$ and a purely power-law function for large $p_{\rm T}$ values. A finite degree of deviation from the equilibrium statistical description of identified particle $p_{\rm T}$ spectra has already been observed by experiments at RHIC \cite{star-prc75,phenix-prc83} and LHC \cite{alice1,alice2,alice3,cms}.  Contrary to a thermalized system, where the $\langle p_{T} \rangle$ is associated with the temperature of the hadronizing matter, one fails to make such a connection in case of systems which are far from thermal equilibrium. In the latter systems, either the temperature fluctuates event by event or within the same event \cite{Bhattacharyya:2015nwa}. This creates room for possible description of the $p_{\rm T}$ spectra in high-energy hadronic and nuclear collisions, using the non-extensive Tsallis statistics \cite{Tsallis:1987eu,Tsallis:2008mc,book}. A thermodynamically consistent  non-extensive distribution function is given by \cite{Cleymans:2011in} 
    \begin{equation}
\label{eq3}
f(m_T) =  C_q \left[1+{(q-1)}{\frac{m_T}{T}}\right]^{-\frac{1}{q-1}} .
\end{equation}
  Here, $m_{\rm T}$ is the transverse mass and $q$ is called the non-extensive parameter- a measure of degree of deviation from equilibrium. Eqs. \ref{eq2} and \ref{eq3} are related through the following transformations for large values of $p_{T}$:
  \begin{equation}
  n= \frac{1}{q-1}, \mathrm{and} ~~~~ p_0 = \frac{T}{q-1}.
  \label{eq4}
  \end{equation} 

\par
 In the limit $q \rightarrow 1$, one recovers the standard Boltzmann-Gibbs distribution (Eq. \ref{eq1}) from the Tsallis distribution  (Eq. \ref{eq5}). Here the effective kinetic freeze-out temperature ($T_{\rm exp}$) as obtained from the inverse slope of the $p_{\rm T}$-spectra using Eq. \ref{eq1}, is related to the Tsallis temperature ($T$) by
 \begin{equation}
  T_{exp} \simeq a + b.T,
  \label{eq4.1}
  \end{equation} 
where $a= 0.31-0.654~q +0.354~q^2$ and $b= 27.35 -55~q+29.07~q^2$ and are obtained numerically for distributions with same mean transverse momentum, $<p_{\rm T}>$, as discussed in Ref. \cite{Rybczynski:2014cha} for $p+p$ collisions in the low-$p_{\rm T}$ ($0.15 < p_{\rm T} < 0.6 $ GeV/c) regime.
     
Tsallis statistics is used widely to explain the particle spectra in high energy collisions \cite{Mishra:2013qoa,Sett:2015lja,Bhattacharyya:2015nwa,Bhattacharyya:2015hya,Zheng:2015gaa,Tang:2008ud,De:2014dna} starting from elementary $e^++e^-$, hadronic and heavy-ion collisions \cite{e+e-,Tripathy:2016hlg,Khuntia:2016ikm,R1,R2,R3,ijmpa,plbwilk,marques,STAR,PHENIX1,PHENIX2,ALICE_charged,ALICE_piplus,CMS1,CMS2,ATLAS,ALICE_PbPb}. The produced particles from the collisions carry the information about collision dynamics and the subsequent space-time evolution till the occurrence of the final freeze-out. The evolution of the partonic system created in high energy experiments is generally believed to be best described by hydrodynamics of an almost ideal fluid \cite{hydro}. This approach gives a fair description of data on the transverse momentum spectra of hadrons, which are treated as one of the important tools to understand the production dynamics of particles in high-energy collisions. The systematic analysis with the help of an appropriate model or approach guides us to understand various thermodynamical as well as hydrodynamical properties of the fireball at different stages of its evolution. The integrated yields of various hadronic species  at different center of mass energies are used in the present work.  The corresponding freeze-out parameters for each hadronic species at the time of their freeze-out can be obtained from the analysis of their respective transverse momentum distributions. Different forms of the invariant yields using Tsallis distribution are available in the literature\cite{Cleymans:2015lxa,Li:2015jpa,Bhattacharyya:2015hya}. In the present work we have used all of these forms to study the temperature ($T$), chemical potential ($\mu$), radial flow ($\beta$), volume (V) and non-extensive parameter ($q$). It should be mentioned here that the parameter $V$ is not necessarily related to the volume one obtains from HBT like experimental measurements. Further, we  study the mass dependence of these parameters, which are obtained by analyzing invariant transverse momentum spectra. For the present analysis, we have used the data of $p+p$ and A+A collisions of different experiments at RHIC and LHC.
We observe a clear mass dependence of the above parameters, and the behaviour is found to be consistent from most peripheral A+A collisions to p+p collisions. The obtained thermodynamic parameters in $p+p$ collisions are similar to those extracted for most peripheral A+A collisions at the same centre of mass energies. This indicates a thermodynamical  similarity between both the systems at a given collision energy.

In the heavy ion collision, the interaction volume of fireball decreases from most central to most peripheral collisions. So the  number of  participant nucleons also decreases from most central collisions to most peripheral collisions depending on the interaction volume. The system having more participant will quickly reach  the equilibrium because of large number of binary collisions by rescattering of partons/hadrons as can be the case in central A+A collisions. But in case of peripheral collisions due to smaller number of participant the system will be away from equilibrium for a while as compared to central collisions. Such a non equilibrium system is better described by Tsallis non-extensive statistics, giving information about the various thermodynamic parameters of the system. 

The paper is organized as follows. In section 2,  we present three forms of Tsallis distribution functions.
Firstly, we discuss invariant yields with and without chemical potential in the Tsallis function. Then we show the Tsallis form of invariant yields with radial flow which is introduced analytically in one of our recent works \cite{Bhattacharyya:2015hya}. In section 3, results and discussions are made. Lastly, we conclude our findings in section 4.   

\section{Non-extensivity and Transverse momentum spectra}
\label{sec:1}
In the following sections, we discuss the transverse momentum spectra of identified particles ($\pi^{+}$, $K^{+}$, p and their antiparticle) produced in RHIC and LHC experiments using different forms of invariant yields using Tsallis non-extensive statistics.

\subsection{Non-extensive statistics without radial flow}
\label{sec:2}
The Tsallis distribution function at mid-rapidity, with finite chemical potential and without radial flow \cite{Cleymans:2015lxa} is given by,

\begin{eqnarray}
\label{eq5}
\left.\frac{1}{p_T}\frac{d^2N}{dp_Tdy}\right|_{y=0} = \frac{gVm_T}{(2\pi)^2}
\left[1+{(q-1)}{\frac{m_T-\mu}{T}}\right]^{-\frac{q}{q-1}}
\end{eqnarray}
 where,  $~ m_T$  is the transverse mass of a particle given by $\sqrt{p_T ^2 + m^2}$, $g$ is the degeneracy and $\mu$ is the chemical potential of the system. In view of higher center of mass energies, where $\mu\simeq 0$, the transverse momentum distribution function~\cite{Li:2015jpa} becomes:
\begin{eqnarray}
\label{eq6}
\left.\frac{1}{p_T}\frac{d^2N}{dp_Tdy}\right|_{y=0} = \frac{gVm_T}{(2\pi)^2}
\left[1+{(q-1)}{\frac{m_T}{T}}\right]^{-\frac{q}{q-1}}.
\end{eqnarray}
 
 \subsection{Non-extensive statistics with radial flow}
The value of the non-extensive parameter $q$ for high energy collisions is  $ 1 \leq q \leq 1.2$ \cite{beck}. To study the order of deviation of the $p_{\rm T}$-spectra from a equilibrium Boltzmann distribution, the Tsallis distribution function has been expanded in a Taylor series in view of $(q-1)$ being very small, after successful inclusion of radial flow in a relativistic scenario. The details of the method are described in Ref.~\cite{Bhattacharyya:2015hya}. The functional form of the distribution up to first order in $(q-1)$ is given by
\bea
\label{eq7}
\frac{1}{2\pi p_T}\frac{dN}{dp_Tdy} &&= \frac{gV}{(2\pi)^3} \nn\\ &&\biggl\{ 2 T [ r I_0(s) K_1(r) - s I_1(s) K_0(r) ]- \nn\\ && (q-1) T r^2  I_0(s) [K_0(r)+K_2(r)]+ \nn\\ && 4(q-1)~T rs I_1(s) K_1(r)-(q-1)Ts^2 K_0(r) \nn\\ &&[I_0(s)+I_2(s)]+ \frac{(q-1)}{4}T r^3 I_0(s) \nn\\ && [K_3(r)+3K_1(r)]- \nn\\ 
&&\frac{3(q-1)}{2} T r^2 s [K_2(r)+K_0(r)] I_1(s)+ \nn\\ && \frac{3(q-1)}{2} T s^2 r [I_0(s)+I_2(s)] K_1(r) \nn\\
&&-\left.\frac{(q-1)}{4}T s^3 [I_3(s)+3I_1(s)] K_0(r)\right\}
\eea
where,$~r\equiv\frac{\gamma m_T}{T} , ~s\equiv\frac{\gamma \beta p_T}{T}.$$~I_n(s)$ and $K_n(r)$ are  the modified Bessel functions of the first
and second kind. There are four parameters involved namely $V$,  $T$,  $\beta$ and $q$; where $V$ is the volume, $T$ 
is the Tsallis temperature, $\beta$ is the radial flow velocity and $q$ is the Tsallis non-extensive parameter.

We use Eq.~\ref{eq7} to fit the particle spectra  of identified particles in heavy ion collisions to study the radial flow parameter.

\section{Results and Discussion} 
It is expected that the number of binary collisions in a system with smaller number of participant nucleons is quite low. Hence, the probability of mutual interaction (resulting in momentum transfer) between the system quanta- partons or hadrons, becomes less for the systems with small value of participating nucleons. This makes the system to stay away from a possible thermal equilibrium. On the other hand, an appreciable increment in the number of binary collisions is observed in a system possessing a large number of participant nucleons and consequently the system can reach quickly to its thermal equilibrium or in the close vicinity of it. It has been shown in Ref.\cite{Ristea:2014qca}, that the non-extensivity parameter ($q$) is close to 1 in central Au+Au collisions and increases towards peripheral collisions. A higher value of $q$ reflects that the system is away from thermal equilibrium. These results suggest that the degree of non-equilibrium is higher in peripheral collisions. Such systems are best described by Tsallis distributions for example the most peripheral A+A and p+p systems. We perform the fitting of $p_{\rm T}$ spectra for $p_{\rm T}<$  3 GeV/c around mid rapidity at  $\sqrt{s_{\rm NN}}$ = 200 GeV and  $\sqrt{s_{\rm NN}}$ = 2.76 TeV. Keeping all the parameters free, we try to fit the spectra with Eqns.~\ref{eq5}, \ref{eq6} and \ref{eq7} independently for different particles like $\pi^{+}$, $K^{+}$, p and their antiparticle for Pb+Pb and Au+Au most peripheral collisions using TMinuit class available in the ROOT library\cite{root} to get a convergent solution. The same procedure is followed for $p+p$ collisions as well.

We have discussed the present work in three sections. 
Now, Eq.~\ref{eq6} is used to analyze invariant $p_{\rm T}$ spectra of identified particles ($\pi^{+}$, $K^{+}$, p and their antiparticles) for most peripheral  Au+Au collisions at $\sqrt{s_{NN}}$ = 200 GeV and most peripheral Pb+Pb collisions at $\sqrt{s_{\rm NN}}$ = 2.76 TeV along with $p+p$ collisions at both energies.The fitting of  $p_{\rm T}$ spectra of identified particles for Au+Au collisions  at 200 GeV and Pb+Pb collisions at 2.76 TeV are shown in Figs. ~\ref{fig1}  and \ref{fig2} (in Appendix) for  most peripheral collisions. Also, Figs. \ref{fig3} and \ref{fig4}  of  appendix section show the fitting of invariant spectra to $p+p$ collisions at 200 GeV and  2.76 TeV,  respectively for identified particles. The fitting is performed up to the maximum $p_{\rm T}$ of 3 GeV/c for A+A collisions and  2.5 GeV/c for $p+p$ collisions although the $K^{+}$  $p_{\rm T}$ spectra in Au+Au collision is fitted  only up-to 2.0 GeV/c due to unavailability 
of data ~\cite{Adler:2003cb}.  The fittings of Tsallis function with $\mu =0$ seem to be better for peripheral heavy-ion collisions. The $\chi^{2}/ndf$ obtained by minimization in TMinuit are shown in Figs. ~\ref{fig5} and ~\ref{fig6}.

 \bef[H]
\begin{center}
\includegraphics[scale=0.35]{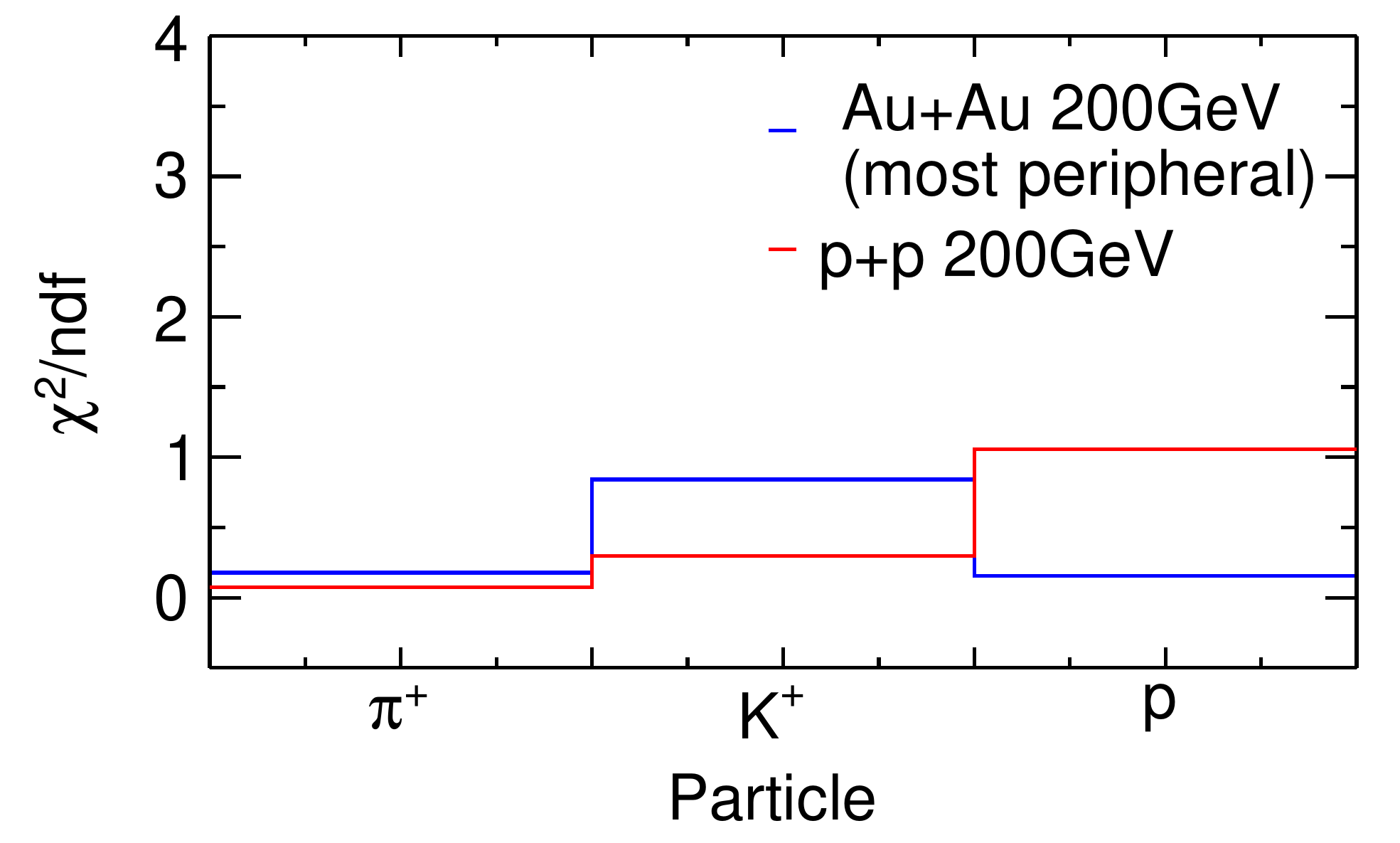} 
\caption{(color online) $\chi^{2}/ndf$ values are shown for most peripheral Au+Au and $p+p$ collisions at $\sqrt{\mathrm{s}_{NN}}$= 200 GeV using Eq.~\ref{eq6} as a fit function.} 
\label{fig5}
\end{center}
\eef

 \bef[H]
 \begin{center} 
\includegraphics[scale=0.35]{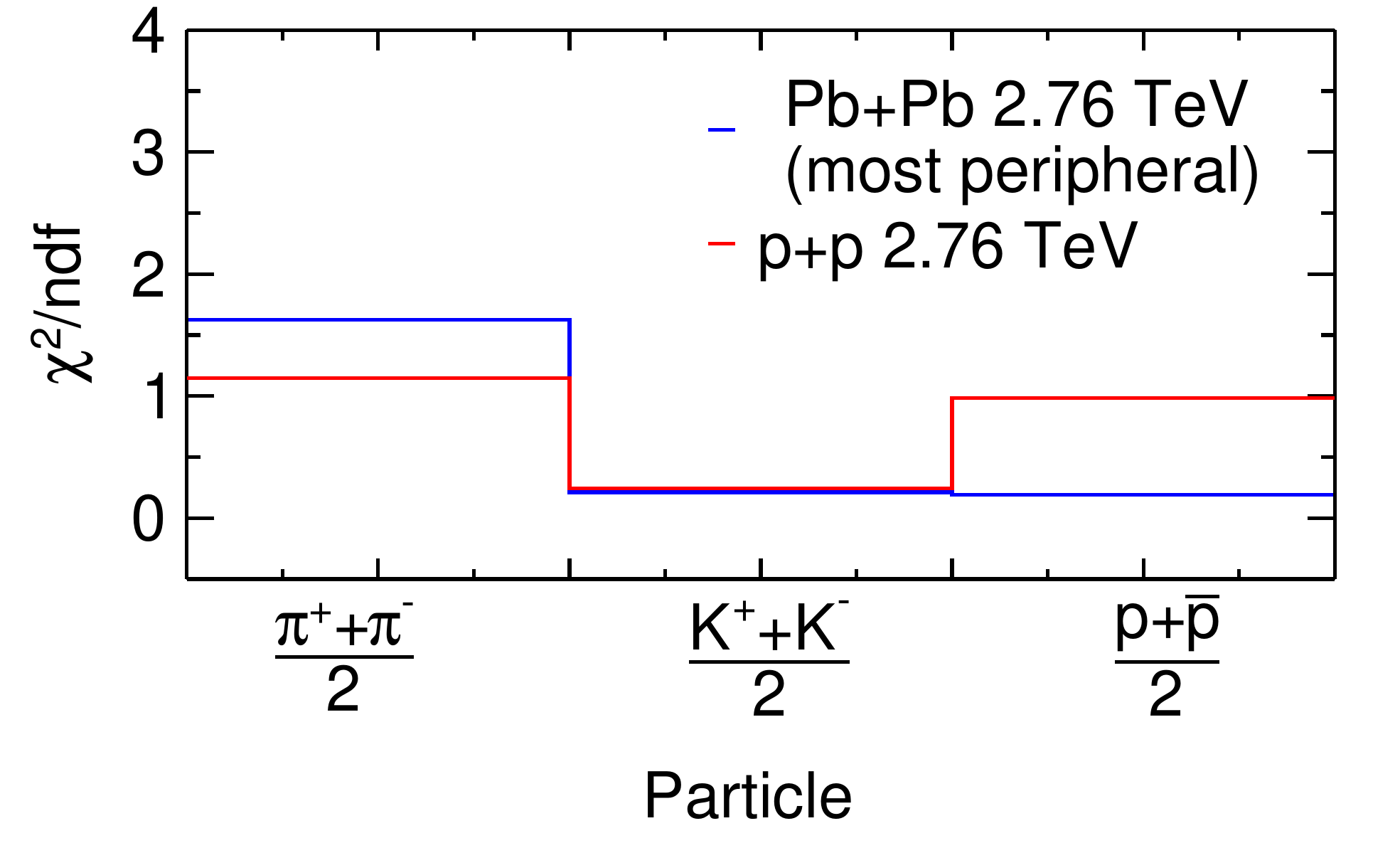} 
 \caption{(color online) $\chi^{2}/ndf$ values are shown for most peripheral Pb+Pb and $p+p$ collisions at $\sqrt{\mathrm{s}_{NN}}$= 2.76 TeV  using Eq.~\ref{eq6} as a fit function.}
\label{fig6}
\end{center}
\eef

 A systematic study of the extracted parameters like volume ($V$),
 temperature ($T$) and Tsallis non-extensive parameter ($q$), is made
 with identified particle masses. The values of extracted parameters are summarized in Table.~\ref{table1} of the appendix. We have calculated the radius of the fireball at freeze-out and plotted as a function of the mass of the identified particles. Figures ~\ref{fig7} and ~\ref{fig8} show the variation of common mass
 dependent  parameters at $\sqrt{s_{\rm NN}}$ = 200 GeV and  $\sqrt{s_{\rm NN}}$ = 2.76 TeV.  We have then  compared the parameters obtained for peripheral heavy-ion collisions with that of $p+p$ collisions at the same centre of mass energies to understand the thermodynamic properties of the produced systems.


\bef[H]	
\begin{center} 
\includegraphics[scale=0.32]{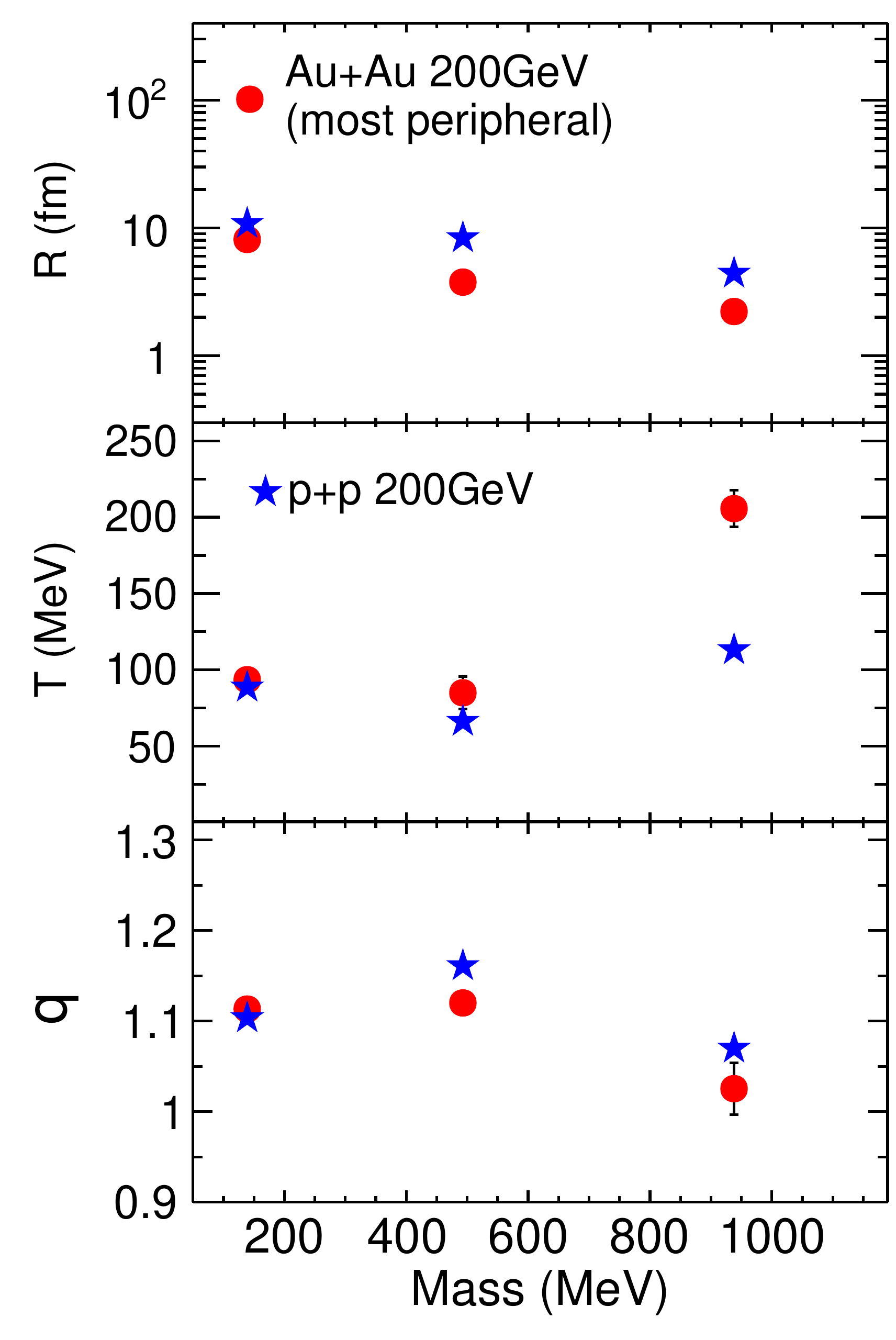}
\caption{(color online) Mass dependence of the thermodynamic parameters using Eq.~\ref{eq6} for  most peripheral Au+Au and $p+p$ collisions at $\sqrt{\mathrm{s}_{NN}}$= 200 GeV.}
\label{fig7}
\end{center}
\eef
  
  It is found that the volume parameter decreases with increase in particle mass. It indicates that different particles have different freeze-out surfaces depending on their mass. This is observed both for $p+p$ and A+A collisions and thus indicates a mass dependent differential freeze-out scenario in high-energy collisions.
In view of the above freeze-out  picture, taking $T$ as the  freeze-out temperature, we observe from Figs. ~\ref{fig7} and ~\ref{fig8} that the freeze-out temperature slightly increases with particle mass. This observation goes inline with the earlier findings of Ref. \cite{Wei:2015oha}, which uses a slightly different approach. It clearly indicates that the heavier particles have higher freeze-out temperature as compared to lighter particles. This observation goes in-line with an intuitive expectation of mass dependent particle freeze-out or differential freeze-out. In literature, one finds various freeze-out scenarios like, single freeze-out \cite{Tang:2008ud}, strange and non-strange particles having two different freeze-outs \cite{Chatterjee:2015fua,Chatterjee:2014lfa} etc. However, our findings from the analysis of high-energy particle spectra with non-extensive statistics reveals a consistent differential freeze-out scenario with particles and their counter parts, the anti-particles freezing out at the same time (characterized by freeze-out temperature, freeze-out volume) from the produced fireballs. The non-extensive measure  i.e., the $q$-parameter obtained from our studies is consistent with its value, $1 < q < 1.2$, both for $p+p$ and A+A collisions. This goes in-line with the expectations of the $q$-values in high-energy collisions \cite{beck}.
 
 \bef[H]
 \begin{center} 
\includegraphics[scale=0.32]{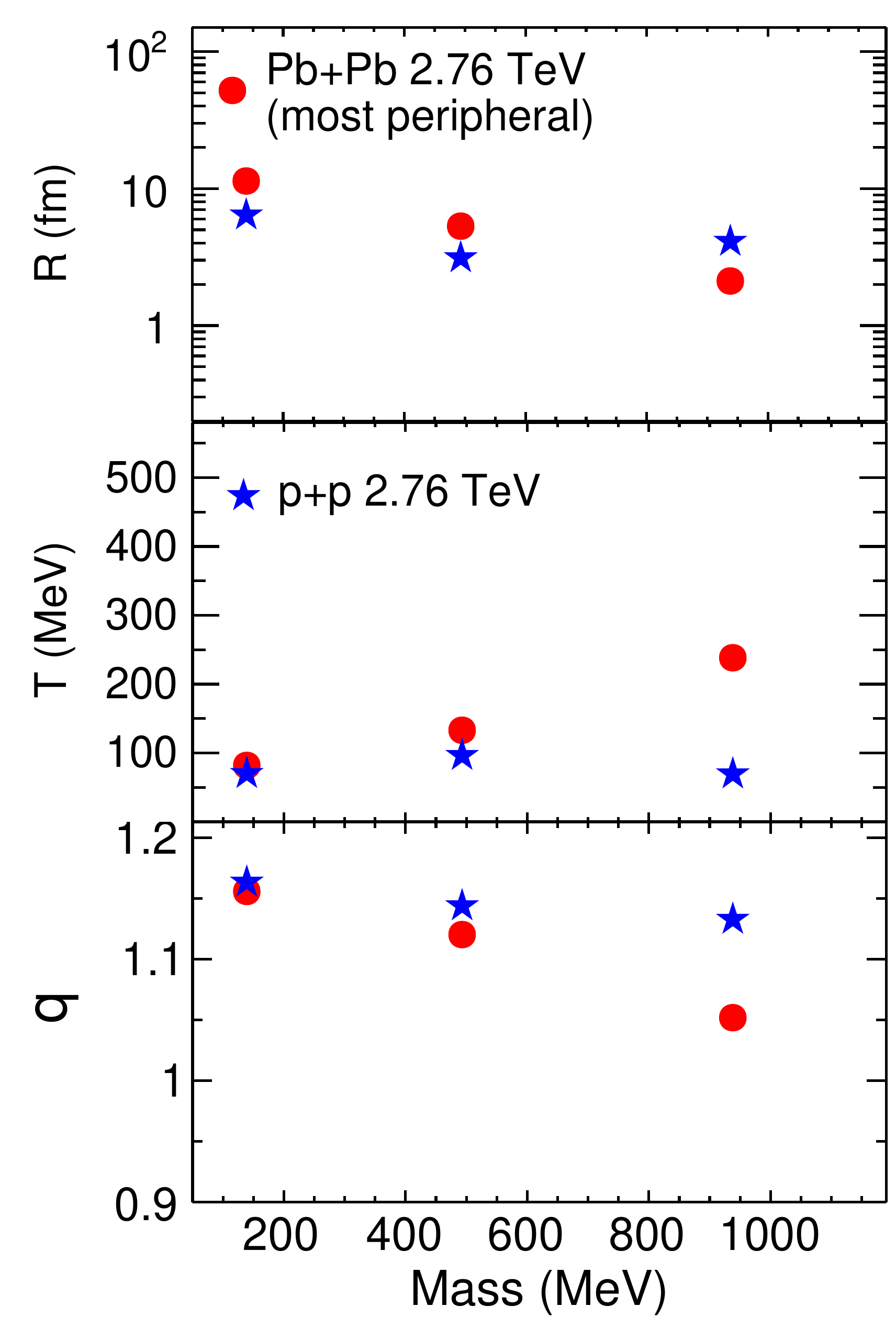}
 \caption{(color online) 
Mass dependence of thermodynamic parameters using Eq.~\ref{eq6} for  most peripheral  Pb+Pb and $p+p$ collisions at $\sqrt{\mathrm{s}_{NN}}$= 2.76 TeV.}
\label{fig8}
\end{center}
\eef

 Also, Figs. ~\ref{fig7} and ~\ref{fig8} indicate a similar mass ordering for $p+p$ systems, which indicates that the system formed in most peripheral  A+A  collisions and $p+p$ collisions are of similar thermodynamic nature. However, it should be noted that the obtained parameters seem to be unphysical {\it i.e.} volume or the radius in $p+p$ collisions seem to be higher than the peripheral A+A collisions at the same energy. This problem is observed to be circumvented, when one adds the radial flow to the distribution. This is seen in the subsequent discussions and in Table-III.
 

Secondly, we perform the fitting of invariant $p_{\rm T}$ spectra of
identified particles using Eq.~\ref{eq5} which contains an additional
parameter as chemical potential ($\mu$). The same fitting  procedure
is used to fit Au+Au most peripheral collisions at
$\sqrt{s_{\rm NN}}$ = 200 GeV for identified particles as well as their antiparticles. The Tsallis form of invariant yields including $\mu$ is used in Fig. ~\ref{fig9}. It fits up to $p_{\rm T}$ $\sim$ 3 GeV/c for most peripheral collisions. The goodness of fit can be seen from $\chi^{2}$/ndf  values shown in Figure ~\ref{fig11}.

\bef[H]
\begin{center}
\includegraphics[scale=0.32]{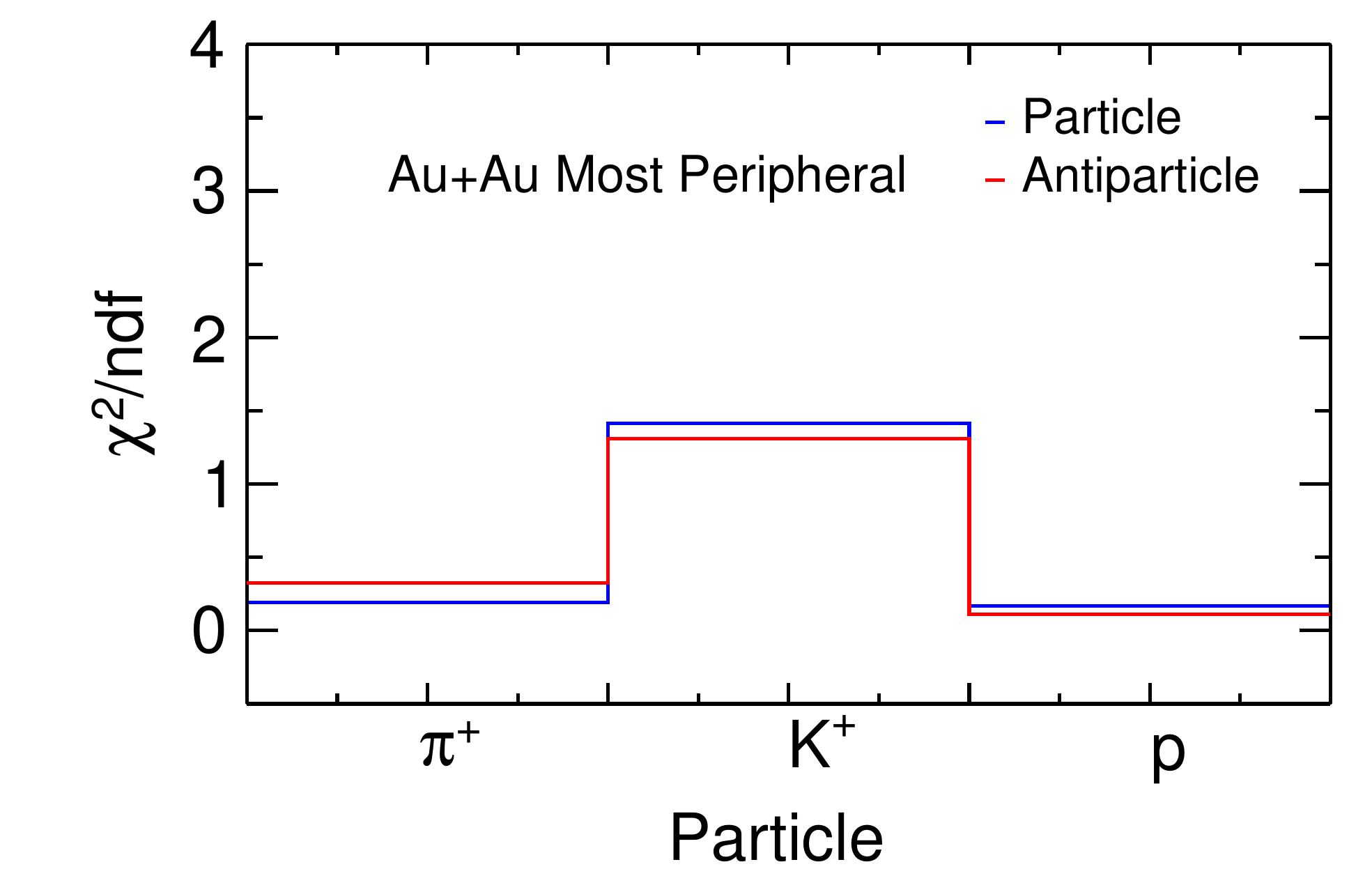}
\newline
\caption{(color online) $\chi^{2}/ndf$ values are shown for most peripheral Au+Au collisions 
at $\sqrt{\mathrm{s}_{NN}}$= 200 GeV using Eq.~\ref{eq5} as a fit function. }

\label{fig11}
\end{center}
\eef

The mass dependence of extracted parameters are shown in the Figure ~\ref{fig12}. Also, the values of the parameters are tabulated in Table.~\ref{table2} of the appendix. In Figure ~\ref{fig12}, it is shown that all the parameters follow the same mass dependent trend as is discussed in the previous cases. Besides, an additional parameter, chemical potential ($\mu$), is shown.

         It is to be mentioned here that, the chemical potential is extracted from the invariant yields at the thermal freeze out, where all the  elastic processes cease. Hence  the obtained chemical potential here is called thermal chemical potential. At this stage one can introduce chemical potentials for all particle species ($\mu_{i}$). Following 
 Ref. \cite{shuryak}, the relationship between chemical potential at kinetic freeze-out and chemical potential at chemical freeze-out is given by                 
        
     \begin{eqnarray}
         \label{eq8}
           \mu_{th} =\mu_{ch}\frac{T_{th}}{T_{ch}} + m(1-\frac{T_{th}}{T_{ch}}) 
         \end{eqnarray}
         
         Where  $(T_{ch} , T_{th})$  and  $(\mu_{ch} , \mu_{th})$  are temperature and chemical potential at chemical and thermal freeze out respectively. Eq.\ref{eq8} clearly
         shows a mass dependency of thermal chemical potential.

            The extracted parameters shown in Fig. ~\ref{fig12}  shows a  consistent picture for particles and antiparticles in case of  most peripheral collisions. It is evident from  Fig. ~\ref{fig12}  that the $\mu$-parameter increases with particle mass.  It shows that particles and antiparticles have same freeze-out conditions.  
It is worth mentioning here that the the parameters $T$ and $\mu$ extracted from identified particle spectra using Eqn. \ref{eq5} are related by \cite{Rybczynski:2014cha}

\begin{equation}
T = T_0  + (q-1) \mu,
\end{equation}
where $T = T_0$ for $\mu=0$. Our extracted parameters using Eqns. \ref{eq5} (i.e. $T, \mu$ and $q$) and \ref{eq6}  (i.e. $T \equiv T_0$) seem to obey the above relation.

\bef[H]
\begin{center}
\includegraphics[scale=0.4]{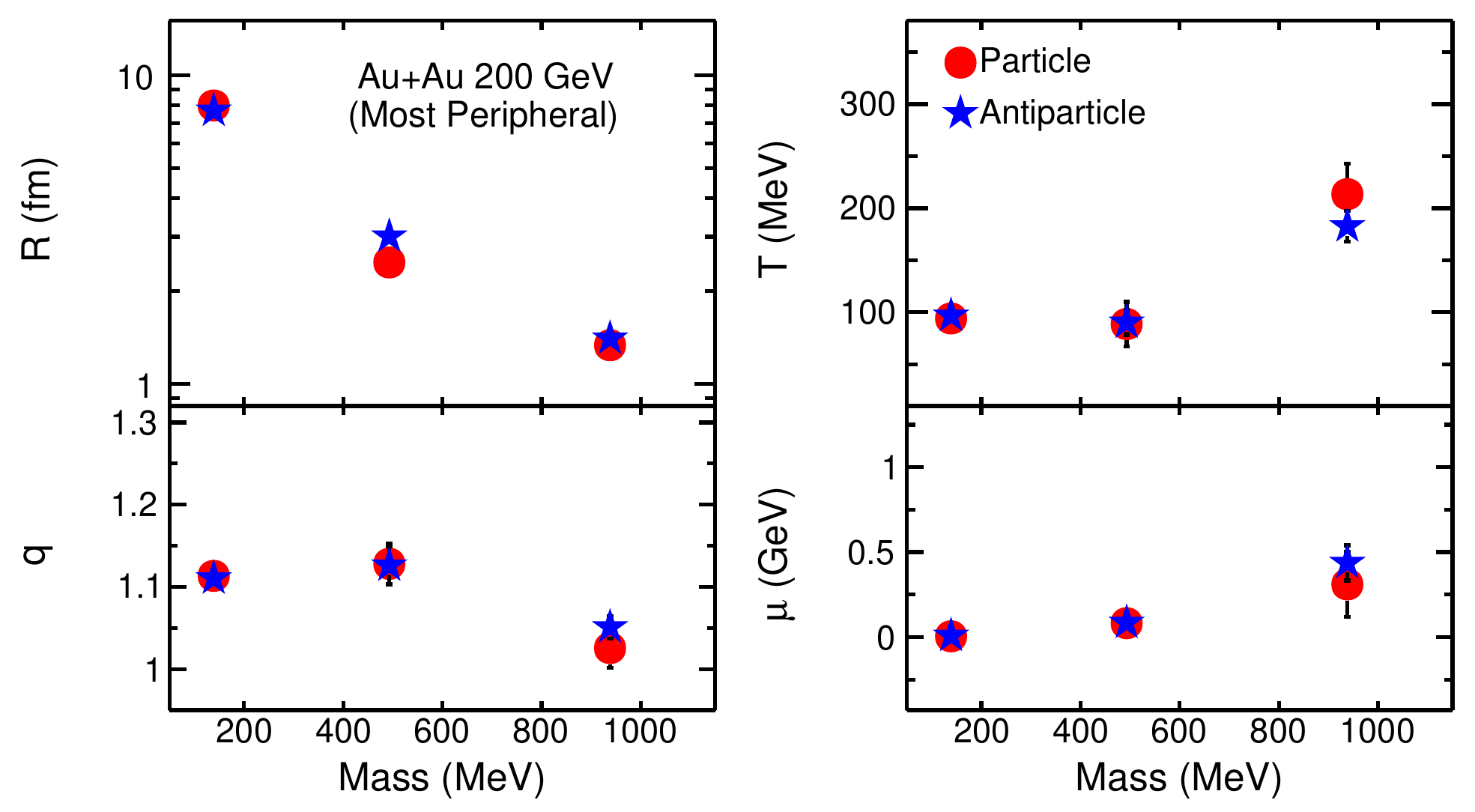}
\newline
\caption{(color online) Mass dependence of extracted thermodynamic parameters using Eq.\ref{eq5}
for different particles and their corresponding antiparticles for most peripheral Au+Au collisions
at $\sqrt{\mathrm{s}_{NN}}$= 200 GeV.}

\label{fig12}
\end{center}
\eef

Finally, the invariant $p_{\rm T}$ spectra of identified particles is
fitted with  Eq.~\ref{eq7} for most  peripheral collisions
of Au+Au at $\sqrt{s_{\rm NN}}$ = 200 GeV and Pb+Pb at $\sqrt{s_{\rm
    NN}}$ = 2.76 TeV. Besides, the fitting of identified particles for
$p+p$ collisions is done for same energies. The fitting results for
Au+Au  collisions at $\sqrt{s_{\rm NN}}$ = 200 GeV  and the results
for Pb+Pb collisions at $\sqrt{s_{\rm NN}}$ = 2.76 TeV are shown in Figs.~\ref{fig14} and \ref{fig15}, respectively.  Also, the fitting results for 
$p+p$ collisions at  $\sqrt{s_{\rm NN}}$ = 200 GeV and $\sqrt{s_{\rm NN}}$ = 2.76 TeV  are shown in  Figs.~\ref{fig16} and \ref{fig17}, respectively in the appendix section.

The fitting is performed up to $p_{\rm T}$ $\sim$ 3 GeV/c in A+A collisions and $p_{\rm T}$ $\sim$ 2.5 GeV/c in $p+p$ collisions, only a slight deviation at higher $p_{\rm T}$ is observed in pion spectra. The values of various parameters obtained from the fits are provided in Table.~\ref{table3}. It is observed from the values of $\chi^{2} /ndf$ (Table.~\ref{table3}) that the performance of the present approach is quite satisfactory, except in case of pion, which can be seen from the Figs.~\ref{fig18} and~\ref{fig19}.

\bef[ht]
\begin{center}
\includegraphics[scale=0.33]{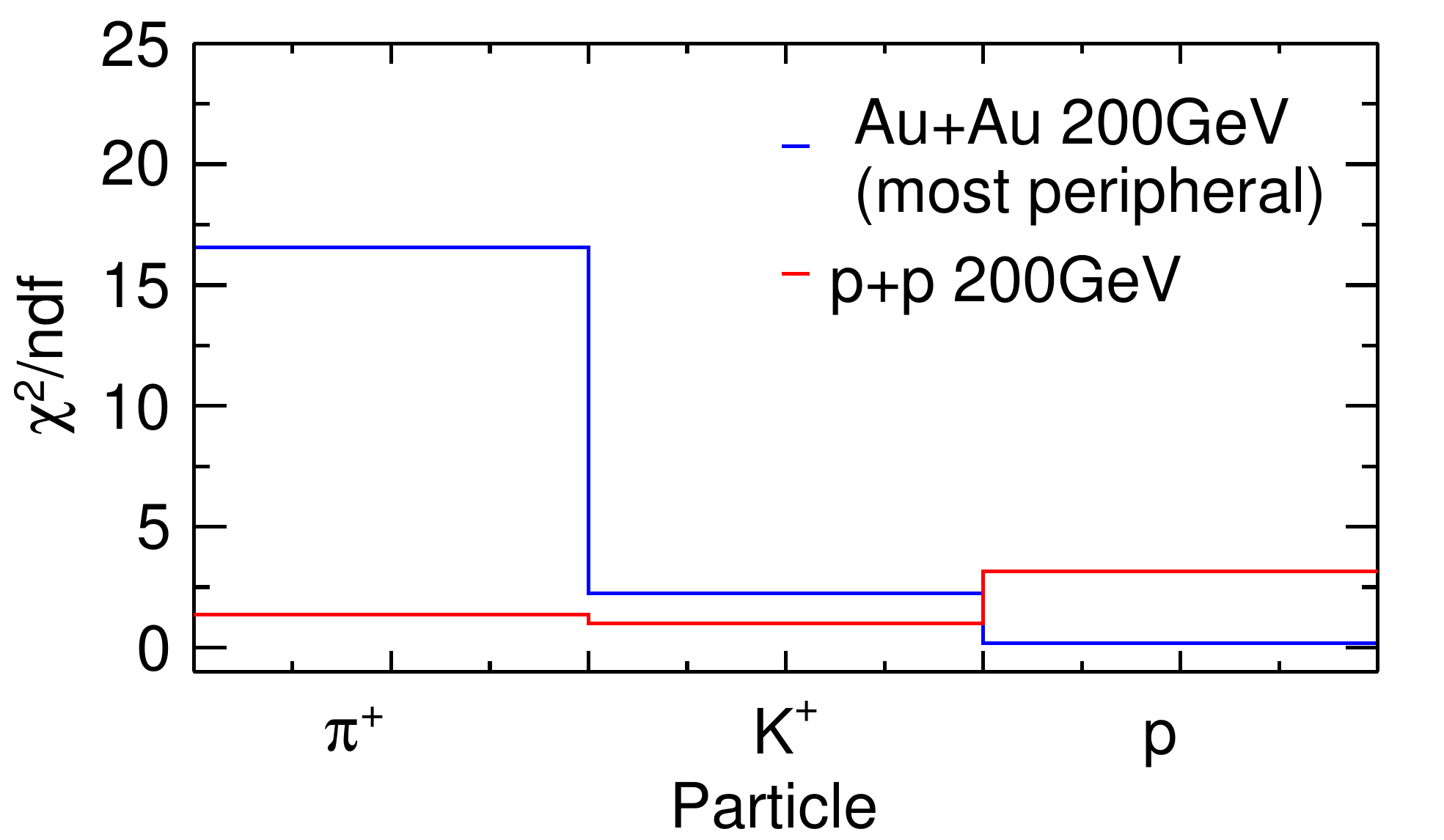}
\newline
\caption{(color online) $\chi^{2}/ndf$ values are shown for most peripheral Au+Au collision and $p+p$ collisions at $\sqrt{\mathrm{s}_{NN}}$= 200 GeV using Eq.~\ref{eq7} as a fit function.}
\label{fig18}
\end{center}
\eef

\bef[ht]
\begin{center}
\includegraphics[scale=0.33]{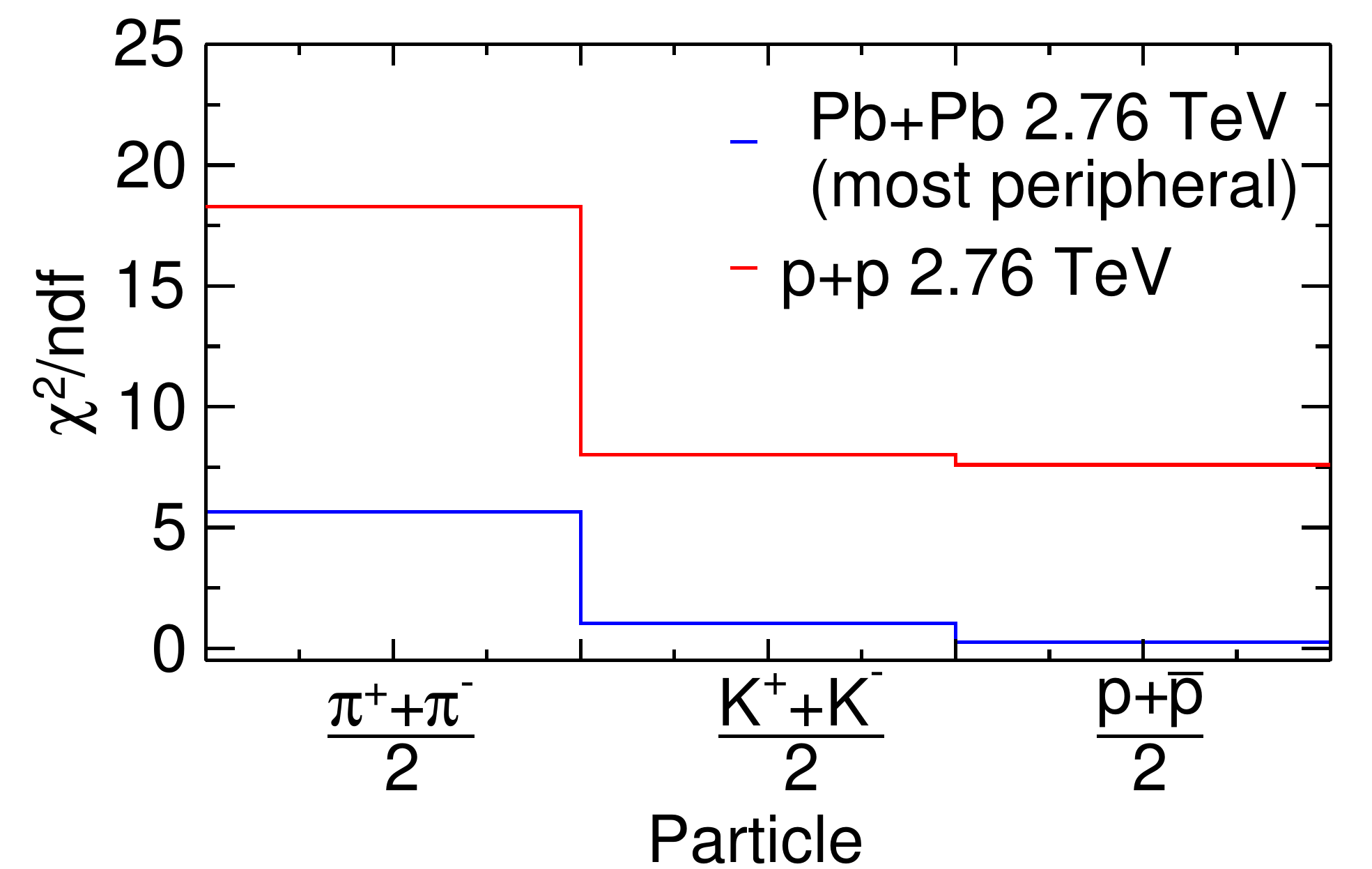}
\newline
\caption{(color online) $\chi^{2}/ndf$ values are shown for most peripheral Pb+Pb collisions and $p+p$ collisions at $\sqrt{\mathrm{s}_{NN}}$= 2.76 TeV using Eq.~\ref{eq7} as a fit function.}
\label{fig19}
\end{center}
\eef

Then we make a mass dependent study of these parameters. Figs.~\ref{fig20} and~\ref{fig21} represent the behaviour of the extracted parameters. 
The parameters like volume, temperature and $q$ show a similar mass dependent behaviour as that of the parameters obtained from the fitting using Eq.~\ref{eq5} and Eq.~\ref{eq6}. 
Here, through Eq.~\ref{eq7}, we have introduced an additional radial flow parameter ($\beta$). 
Figs. ~\ref{fig20} and \ref{fig21} show the decrease of radial flow with increase of particle mass, which is a signature of hydrodynamic evolution of the system created in high energy collisions.
\
The change of temperature parameter with the use of different forms of Tsallis distribution  could be observed in Ref.\cite{Zheng:2015mhz}.   In our analysis we have also observed not only the temperature  but also the other parameters are changing with the use of different forms of Tsallis distributions. For example, It is observed that with the inclusion of radial flow, the value of the $q$-parameter  becomes less, as compared to that observed in the other forms of Tsallis distribution, while describing the $p_{\rm T}$-spectra. This may hint that the non-extensivity is shared by the dynamics of the system\cite{Biyajima:2006mv,biro1,biro2}.   

\bef[H]
\begin{center}
\includegraphics[scale=0.39]{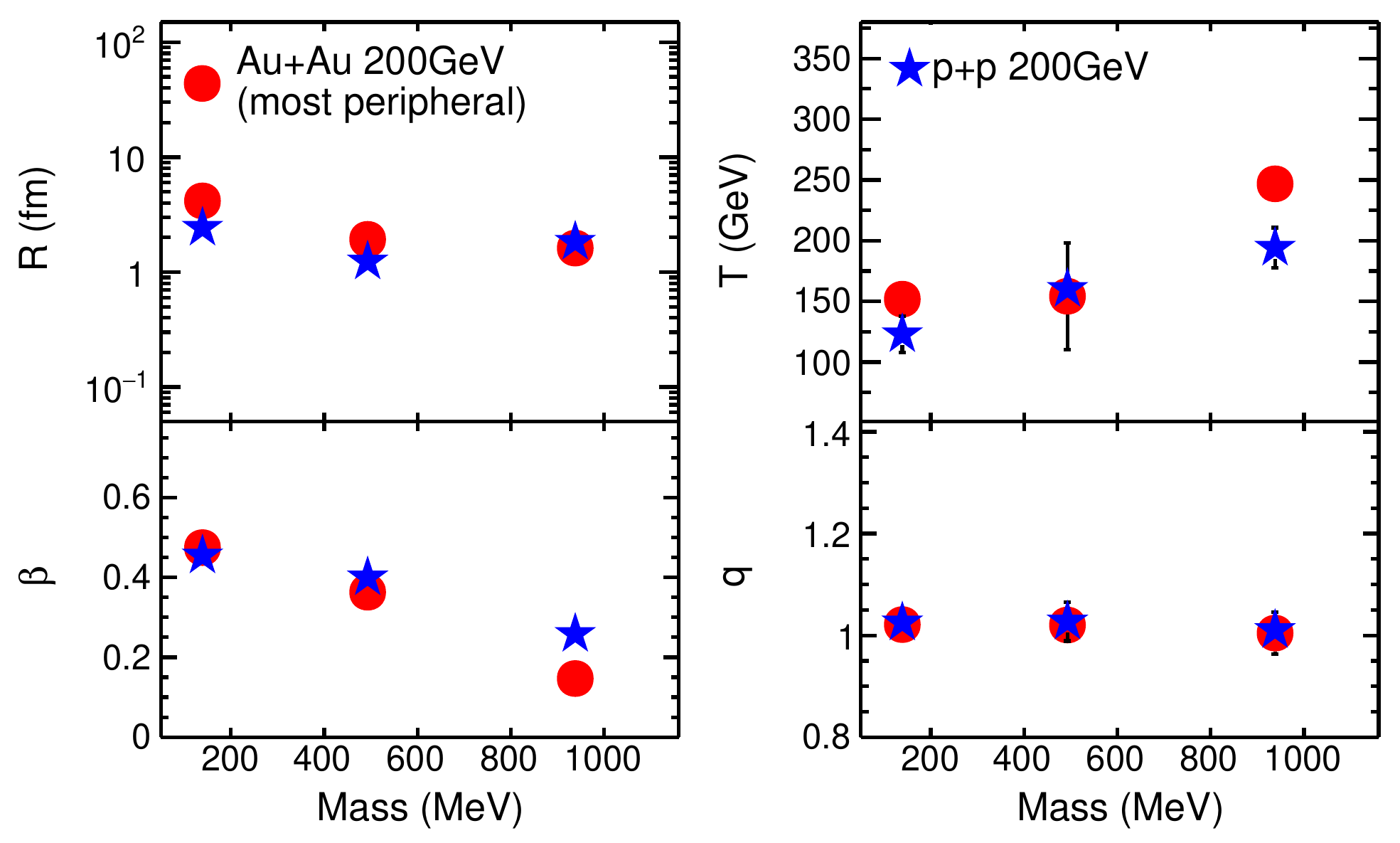}
\newline
\caption{(color online) Mass dependence of extracted thermodynamic parameters of Tsallis distribution including radial flow using Eq.~\ref{eq7}  for most peripheral Au+Au collisions and $p+p$ collisions at $\sqrt{\mathrm{s}_{NN}}$= 200 GeV.}

\label{fig20}
\end{center}
\eef

\bef[H]
\begin{center}
\includegraphics[scale=0.39]{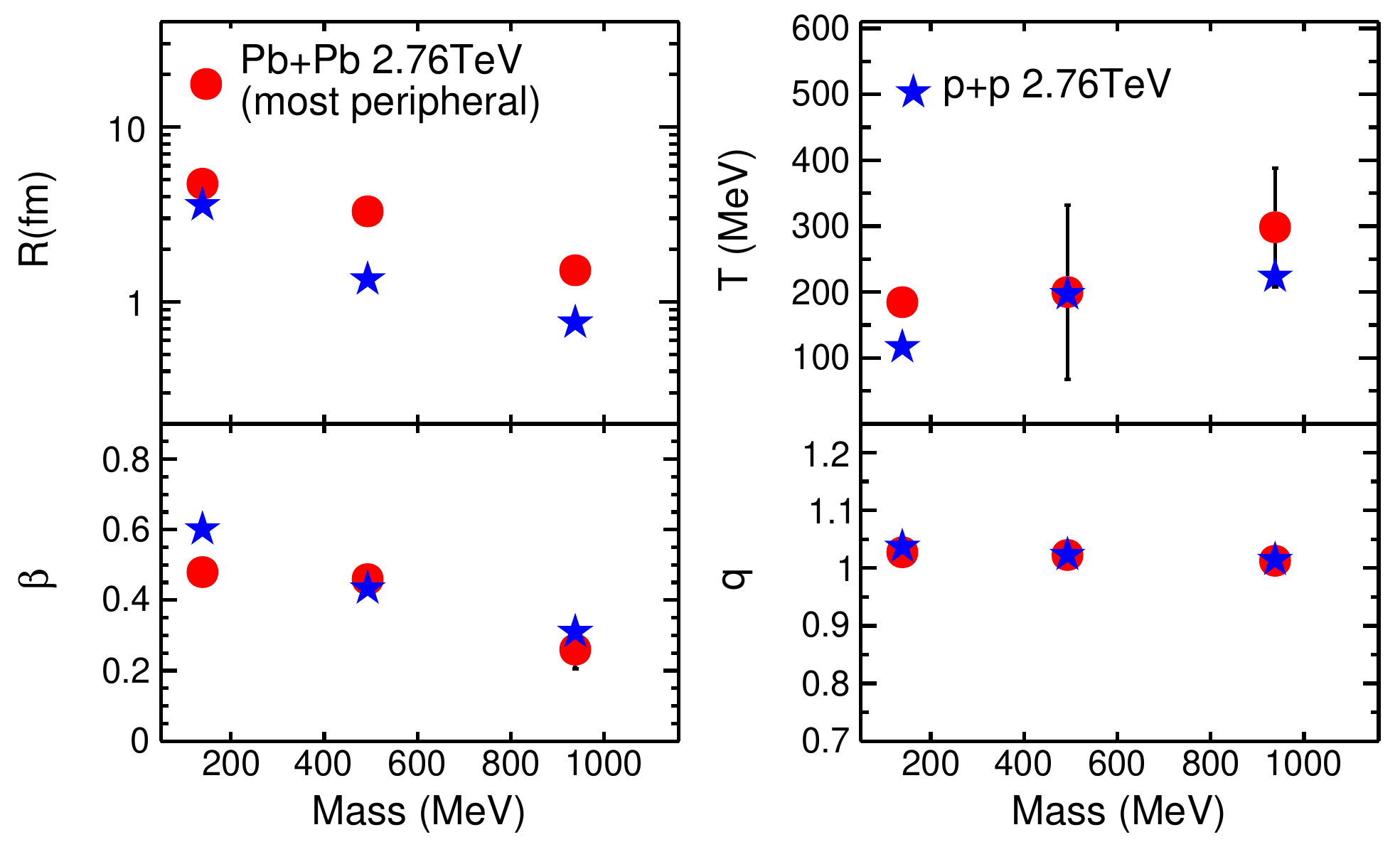}
\newline
\caption{(color online) Mass dependence of extracted thermodynamic parameters of Tsallis distribution including radial flow using Eq.~\ref{eq7} for most peripheral Pb+Pb collisions and $p+p$ collisions at $\sqrt{\mathrm{s}_{NN}}$= 2.76 TeV.
}
\label{fig21}
\end{center}
\eef

\section{Conclusion}
 We present a systematic study of the parameters extracted from transverse momentum spectra of produced identified particles in Au+Au collisions at $\sqrt{\mathrm{s}_{NN}}$= 200 GeV and Pb+Pb collisions at $\sqrt{\mathrm{s}_{NN}}$= 2.76 TeV. Further, the same analysis is done for $p+p$ collisions at both, RHIC and LHC energies. 
 In the present work different forms of invariant yields using  Tsallis distribution are used. Parameters  obtained from three different forms in most peripheral Au+Au collisions at $\sqrt{\mathrm{s}_{NN}}$= 200 GeV and in most peripheral Pb+Pb collisions at  $\sqrt{\mathrm{s}_{NN}}$= 2.76 TeV follow a similar trend. Similar behaviour has been observed in $p+p$ collisions for both the energies. The chemical potential plays an important role in particle production mechanism. We observe that the particles produced at early stage  have higher thermal chemical potential as compared to particles produced at the later stage. The radial flow extracted from the particle spectra of peripheral A+A collisions is similar to $p+p$ collisions at the same center of mass energy. In the discussed low-$p_{\rm T}$ region, the radial flow decreases with increasing mass, which goes in line with hydrodynamic description of the evolution of a fireball created in high-energy collisions. Therefore, the present analysis gives a systematic information of hydrodynamical and thermodynamical evolution of the system, while making a direct comparison with hadronic and peripheral nuclear collisions. We observe a mass dependent differential freeze-out scenario in high-energy hadronic and nuclear collisions, which is also supported by similar observations by other authors \cite{Hai}.

\section*{Acknowledgements}

D.Thakur acknowledges the financial support from U.G.C, New Delhi and S. Tripathy acknowledges the DST-INSPIRE scheme of Govt. of India for financial support.


\section{Appendix}


\begin{table*}

\centering

\caption { ${\chi ^2}/{ndf}$ and different extracted parameters from Tsallis distribution fit(Eq.~\ref{eq6}) to the $p_T$ spectra of 
(a) $\pi ^+$, $K^+$ and $p$ for most peripheral
Au+Au collisions and $p+p$ collisions at $\sqrt{\mathrm{s}_{NN}}$= 200 GeV; (b) $\frac{\pi ^+ + \pi ^- }{2}$, $\frac{K^+ + K^-}{2}$ and $\frac{p + \bar p}{2}$ for most peripheral
Pb+Pb collisions and $\frac{\pi ^+ + \pi ^- }{2}$, $\frac{K^+ + K^-}{2}$ and $\frac{p + \bar p}{2}$ for $p+p$ collisions at $\sqrt{\mathrm{s}_{NN}}$= 2.76 TeV.}


\subfloat [] {

\noindent\begin{tabular}{ |c|c|c|c|c| } 
\hline
\multicolumn{5}{|c|}{\bf{Au+Au 200 GeV (Peripheral)}}\\
\hline
Particle & ${\chi ^2}/{ndf}$ & V $(fm)^3$ & T $(GeV)$ & q \\
\hline
$\pi ^+$       & 0.178 & 2242.35 $\pm$ 18.23  & 0.093 $\pm$ 0.003 & 1.113 $\pm$ 0.004 \\
$K^+$         & 0.842 & 223.358 $\pm$ 0.010    & 0.085 $\pm$ 0.011 & 1.120 $\pm$ 0.002 \\
$p$              & 0.157 & 45.62 $\pm$ 2.45   & 0.205 $\pm$ 0.012   & 1.025 $\pm$ 0.029 \\

\hline
\multicolumn{5}{|c|}{\bf{$p+p$ 200 GeV}}\\
\hline
$\pi ^+$        & 0.073 & 5293.1 $\pm$ 23.3   & 0.088 $\pm$ 0.004 & 1.103 $\pm$ 0.005  \\
$K^+$          & 0.297 & 2443.7 $\pm$ 33.7   & 0.066 $\pm$ 0.002 & 1.161 $\pm$ 0.009  \\
$p$              & 1.055 & 360.5 $\pm$ 9.6    & 0.113 $\pm$ 0.004 & 1.070 $\pm$ 0.000 \\

\hline
\end{tabular}

}





\subfloat [] {

\begin{tabular}{ |c|c|c|c|c| } 
\hline
\multicolumn{5}{|c|}{\bf{Pb+Pb 2.76 TeV (Peripheral)}}\\
\hline
Particle& ${\chi ^2}/{ndf}$ & V $(fm)^3$ & T $(GeV)$ & q \\
\hline
$\frac{\pi ^+ + \pi ^- }{2}$  & 1.626 & 6134.30 $\pm$ 12.61 & 0.082 $\pm$ 0.001 & 1.156 $\pm$ 0.002 \\
$\frac{K^+ + K^-}{2}$        & 0.211 & 629.69 $\pm$ 14.80 & 0.133 $\pm$ 0.003 & 1.120 $\pm$ 0.004 \\
$\frac{p + \bar p}{2}$        & 0.193 & 39.623 $\pm$ 1.052 & 0.238 $\pm$ 0.012 & 1.052 $\pm$ 0.009 \\

\hline
\multicolumn{5}{|c|}{\bf{$p+p$ 2.76 TeV}}\\
\hline
$\frac{\pi ^+ + \pi ^- }{2}$ & 0.985 & 1102.46 $\pm$ 6.82 & 0.070 $\pm$ 0.001 & 1.163 $\pm$ 0.001 \\
$\frac{K^+ + K^-}{2}$       & 0.094 & 127.13 $\pm$ 1.33 & 0.096 $\pm$ 0.005 & 1.144 $\pm$ 0.005 \\
$\frac{p + \bar p}{2}$       & 0.006 & 291.39 $\pm$ 10.62 & 0.069 $\pm$ 0.003 & 1.133 $\pm$ 0.002 \\

\hline
\end{tabular}

}


\label{table1}
\end{table*}

\begin{table*}

\centering
\caption{${\chi ^2}/{ndf}$ and different extracted parameters from Tsallis distribution fit(Eq.~\ref{eq5}) to the $p_T$ spectra of 
$\pi ^\pm $, $K^\pm $, $p$ and $\bar p$  for most peripheral
Au+Au collisions at $\sqrt{\mathrm{s}_{NN}}$= 200 GeV.}

\begin{tabular}{ |c|c|c|c|c|c| } 
\hline
\multicolumn{6}{|c|}{\bf{Au+Au 200 GeV (Peripheral)}}\\
\hline
Particle& ${\chi ^2}/{ndf}$ & V $(fm)^3$ & T $(GeV)$ & q & $\mu$ $(GeV)$ \\
\hline
$\pi ^+$  & 0.191 & 2125.93 $\pm$ 56.06 & 0.094   $\pm$ 0.004 & 1.113 $\pm$ 0.004 & 0.0050 $\pm$ 0.0003 \\
$\pi ^-$   & 0.325 & 1895.78 $\pm$ 275.34 & 0.097 $\pm$ 0.004 & 1.110 $\pm$ 0.004 & 0.0080 $\pm$ 0.0003 \\
$K^+$    & 1.415 & 63.64 $\pm$ 12.60 & 0.089   $\pm$ 0.021 & 1.128 $\pm$ 0.025 & 0.086 $\pm$ 0.009 \\
$K^-$     & 1.309 & 113.82 $\pm$ 68.83 & 0.090   $\pm$ 0.012 & 1.126 $\pm$ 0.023 & 0.085 $\pm$ 0.003 \\
$p$        & 0.170 & 10.01 $\pm$ 3.71 & 0.213   $\pm$ 0.029 & 1.026 $\pm$ 0.024 & 0.310 $\pm$ 0.190\\
$\bar p$ & 0.111 & 11.62 $\pm$ 3.48 & 0.183   $\pm$ 0.015 & 1.051 $\pm$ 0.014 & 0.437 $\pm$ 0.103 \\
\hline

\end{tabular}

\label{table2}

\end{table*}

\begin{table*} 
 
\centering

\caption {${\chi ^2}/{ndf}$ and different extracted parameters from Tsallis distribution fit(Eq.~\ref{eq7}) to the $p_T$ spectra of (a) $\pi ^+$, $K^+$ and $p$ for most peripheral
Au+Au collisions and $p+p$ collisions at $\sqrt{\mathrm{s}_{NN}}$= 200 GeV; 
(b) $\frac{\pi ^+ + \pi ^- }{2}$, $\frac{K^+ + K^-}{2}$ and $\frac{p + \bar p}{2}$ for most peripheral
Pb+Pb collisions and $\frac{\pi ^+ + \pi ^- }{2}$, $\frac{K^+ + K^-}{2}$ and $\frac{p + \bar p}{2}$ for $p+p$ collisions at $\sqrt{\mathrm{s}_{NN}}$= 2.76 TeV.}


\subfloat [] {
\begin{tabular}{ |c|c|c|c|c|c| } 
\hline
\multicolumn{6}{|c|}{\bf{Au+Au 200 GeV (Peripheral)}}\\
\hline
Particle & ${\chi ^2}/{ndf}$ & V $(fm)^3$ & T $(GeV)$ & $\beta$  & q \\
\hline
$\pi ^+$       & 16.558   & 302.680 $\pm$ 7.591  & 0.152 $\pm$ 0.001   & 0.475 $\pm$ 0.013 & 1.021   $\pm$ 0.004 \\
$K^+$         & 2.2461   & 30.005 $\pm$ 1.676  & 0.154 $\pm$ 0.044   & 0.363 $\pm$ 0.008 & 1.020   $\pm$ 0.021 \\
$p$             & 0.1825   & 17.731 $\pm$ 1.952  & 0.247 $\pm$ 0.000   & 0.147 $\pm$ 0.018 & 1.004   $\pm$ 0.041 \\

\hline

\multicolumn{6}{|c|}{\bf{$p+p$ 200 GeV}}\\

\hline
$\pi ^+$       & 1.370 & 60.260 $\pm$ 0.506 & 0.123 $\pm$ 0.015 & 0.455 $\pm$ 0.015  & 1.025 $\pm$ 0.015 \\
$K^+$          & 0.996 & 7.914 $\pm$ 0.878 & 0.160 $\pm$ 0.004 & 0.400 $\pm$ 0.013  & 1.027 $\pm$ 0.038 \\
$p$              & 3.144 & 25.970 $\pm$ 0.667 & 0.194 $\pm$ 0.017 & 0.258 $\pm$ 0.019  & 1.010 $\pm$ 0.010 \\

\hline

\end{tabular}
}



\begin{center}
\subfloat [] {

\begin{tabular}{ |c|c|c|c|c|c| } 
\hline
\multicolumn{6}{|c|}{\bf{Pb+Pb 2.76 TeV (Peripheral)}}\\
\hline
Particle & ${\chi ^2}/{ndf}$ & V $(fm)^3$ & T $(GeV)$ & $\beta$ & q \\
\hline
$\frac{\pi ^+ + \pi ^- }{2}$  & 5.657 & 444.714 $\pm$ 7.931 & 0.185 $\pm$ 0.008 & 0.480 $\pm$ 0.028 & 1.027 $\pm$ 0.006 \\
$\frac{K^+ + K^-}{2}$        & 1.025 & 149.038 $\pm$ 11.465 & 0.200 $\pm$ 0.133 & 0.460 $\pm$ 0.021 & 1.023 $\pm$ 0.010 \\
$\frac{p + \bar p}{2}$        & 0.258 & 14.507 $\pm$ 1.107 & 0.299 $\pm$ 0.090 & 0.260 $\pm$ 0.055 & 1.013 $\pm$ 0.024 \\

\hline
\multicolumn{6}{|c|}{\bf{$p+p$ 2.76 TeV}}\\

\hline
$\frac{\pi ^+ + \pi ^- }{2}$  & 18.290 & 191.039 $\pm$ 15.890 & 0.116 $\pm$ 0.003 & 0.601 $\pm$ 0.013 & 1.038 $\pm$ 0.004 \\
$\frac{K^+ + K^-}{2}$         & 8.006 & 10.164 $\pm$ 1.097 & 0.197 $\pm$ 0.005 & 0.433 $\pm$ 0.016 & 1.023 $\pm$ 0.005 \\
$\frac{p + \bar p}{2}$        & 7.596 & 1.839 $\pm$ 0.256 & 0.224 $\pm$ 0.006 & 0.310 $\pm$ 0.014 & 1.014 $\pm$ 0.004 \\

\hline
\end{tabular}

}

\centering
\label{table3}
\end{center}

\end{table*}

\clearpage

\bef[H]
\begin{center}
\includegraphics[scale=0.28]{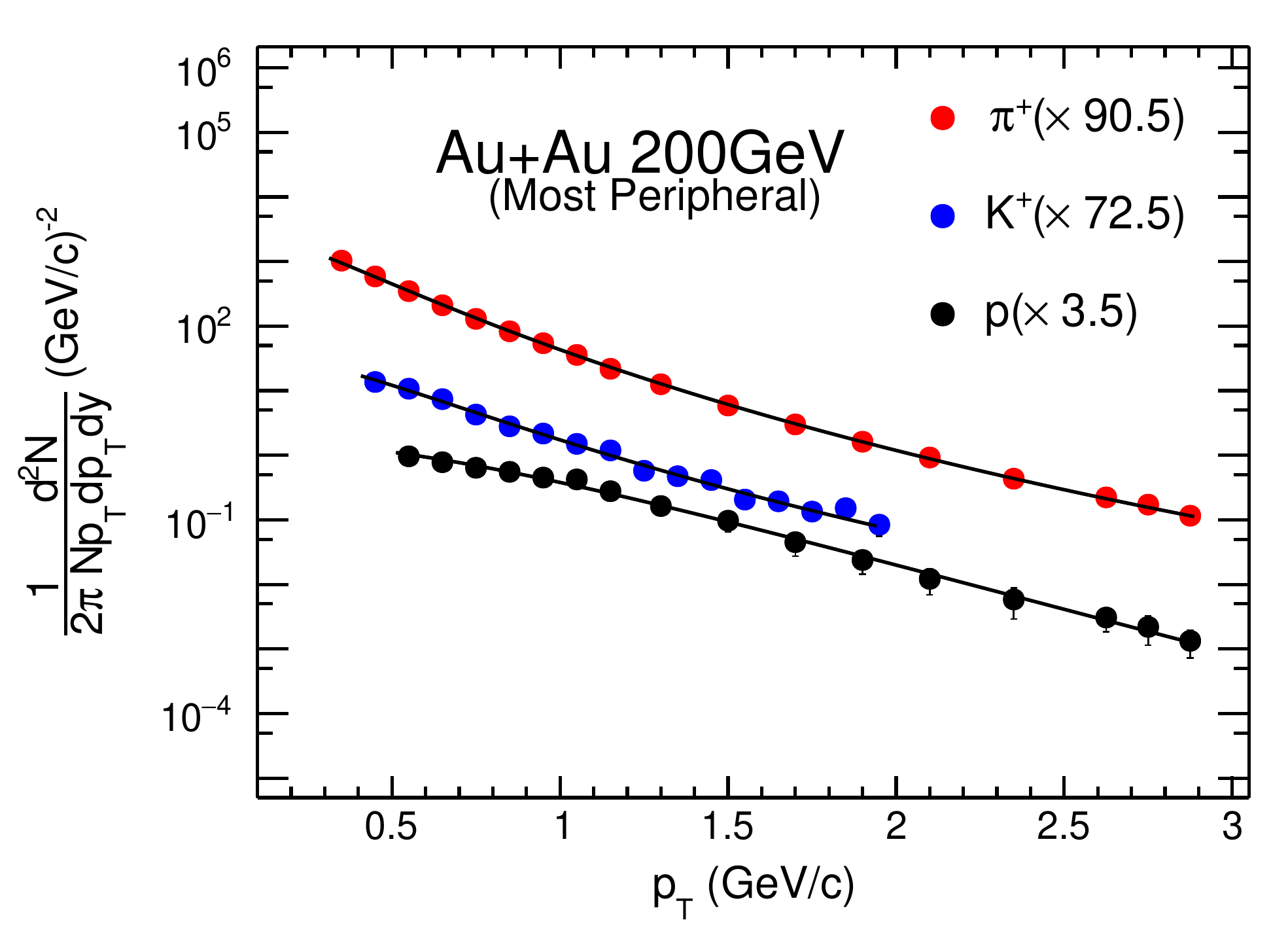}	
\caption {(color online) Fitting of  invariant $p_{\rm T}$ spectra with Tsallis distribution using Eq.~\ref{eq6}  for 
$\pi^{+}$ \cite{Adler:2003cb}, $K^{+}$ \cite{Adler:2003cb},
$p$ \cite{Adler:2003cb}  in most  peripheral Au+Au collisions at $\sqrt{\mathrm{s}_{NN}}$= 200 GeV. 
Symbols represent the data points and line is the fit function.}
\label{fig1}
\end{center}
\eef

\bef[H]
\begin{center}
\includegraphics[scale=0.28]{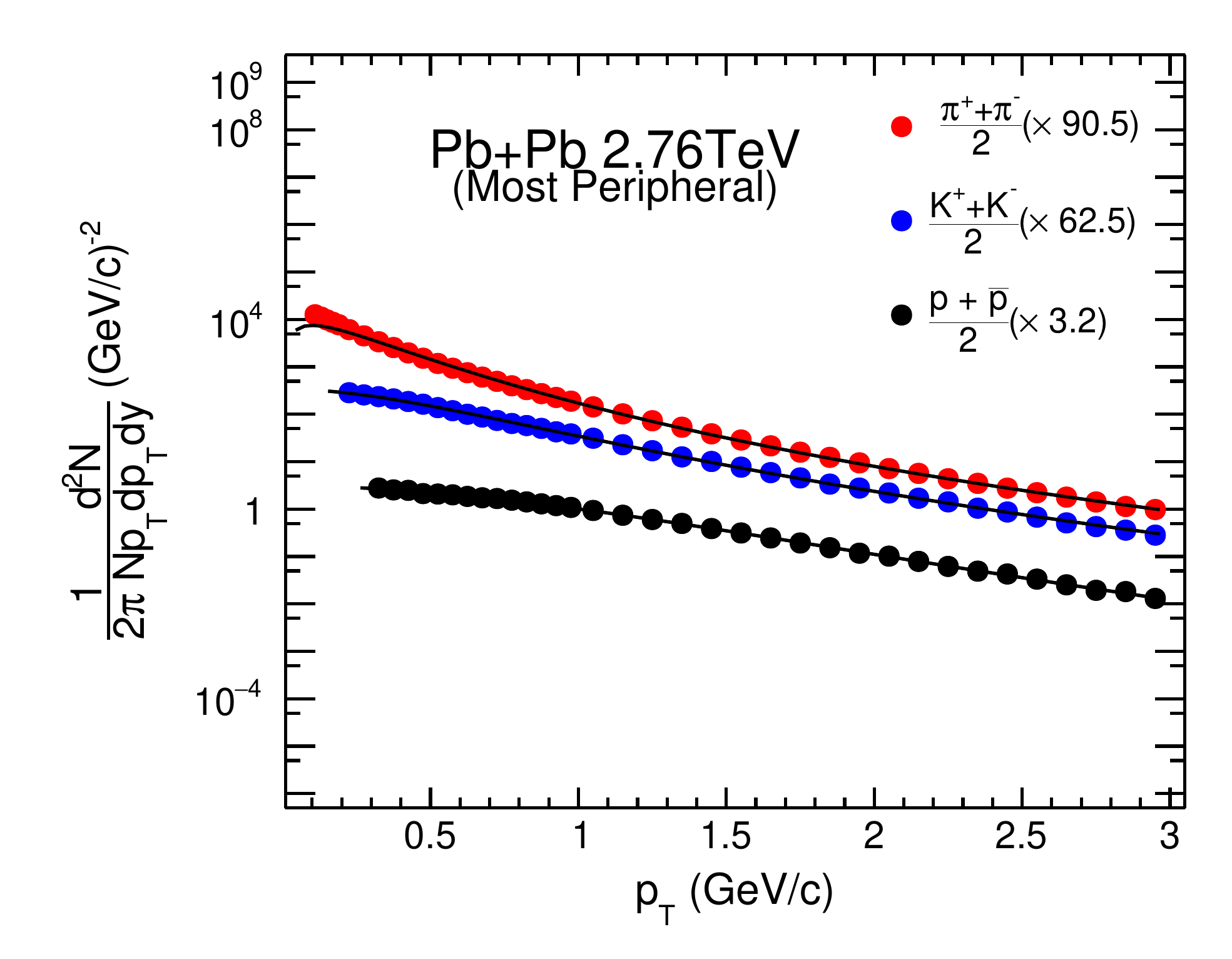}
\caption {(color online) Fitting of  invariant $p_{\rm T}$ spectra with Tsallis distribution using Eq.~\ref{eq6} for  
$\frac{\pi^ {+}+ \pi^{-}}{2}$ \cite{Abelev:2014laa}, $\frac{K ^ {+}+K ^{-}}{2}$ \cite{Abelev:2014laa}, $\frac{p+\bar p}{2}$ \cite{Abelev:2014laa} in most  peripheral Pb+Pb collisions at $\sqrt{\mathrm{s}_{NN}}$=2.76 TeV. Symbols represent the data points and line is the fit function.}
\label{fig2}
\end{center}
\eef

\bef [H]
\begin{center} 
\includegraphics[scale=0.32]{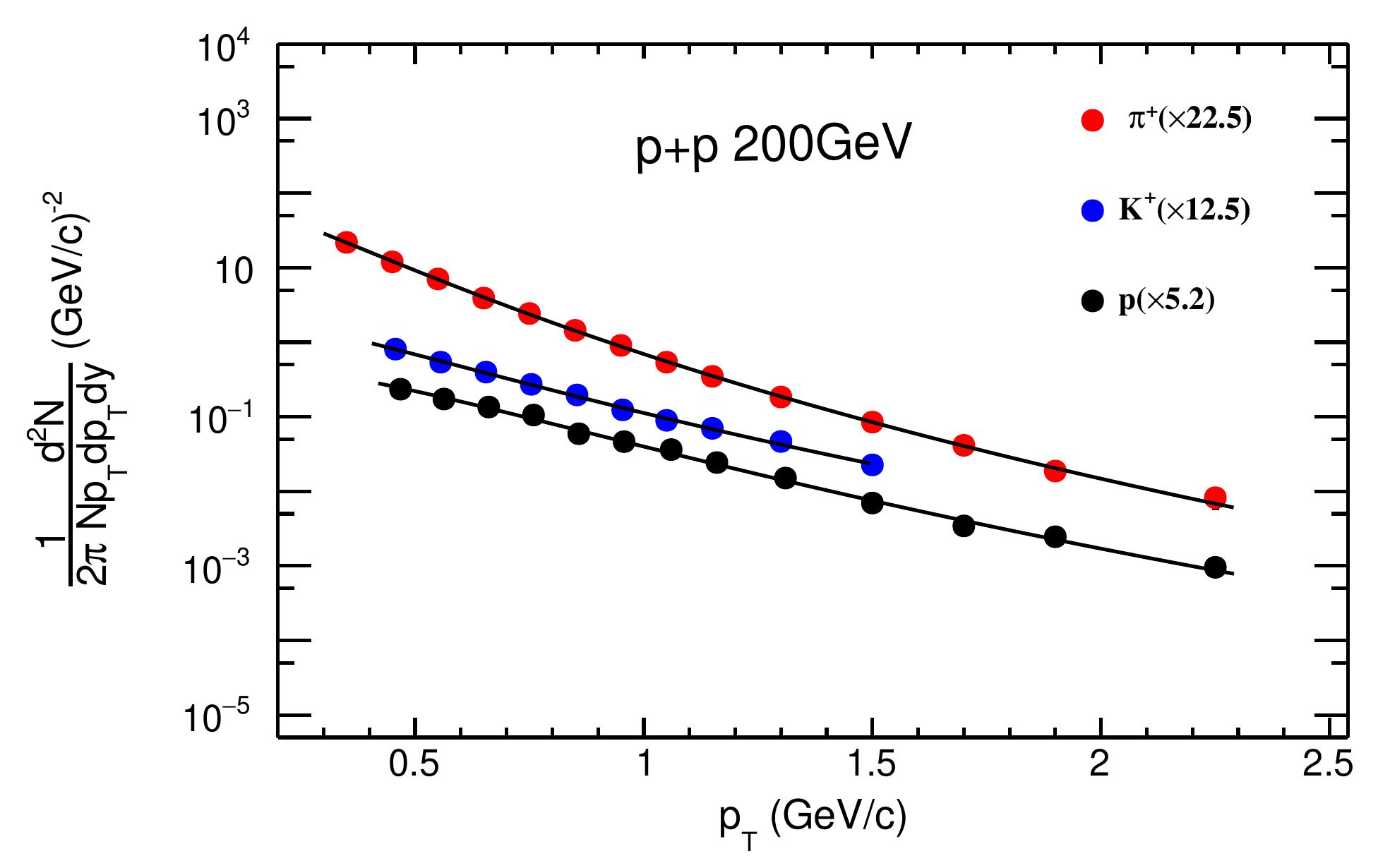}
\caption {(color online)
Fitting of  invariant $p_{\rm T}$ spectra with Tsallis distribution using Eq.~\ref{eq6} for $\pi^{+} $ \cite{Adams:2003qm}, $K^{+}$  \cite{Adams:2003qm}, $p$ \cite{Adams:2003qm}  in $p+p$ collisions at $\sqrt{\mathrm{s}_{NN}}$= 200 GeV. Symbols represent the data points and line is the fit function. } 
\label{fig3}
\end{center}
\eef

\bef[H]
\begin{center}
\includegraphics[scale=0.24]{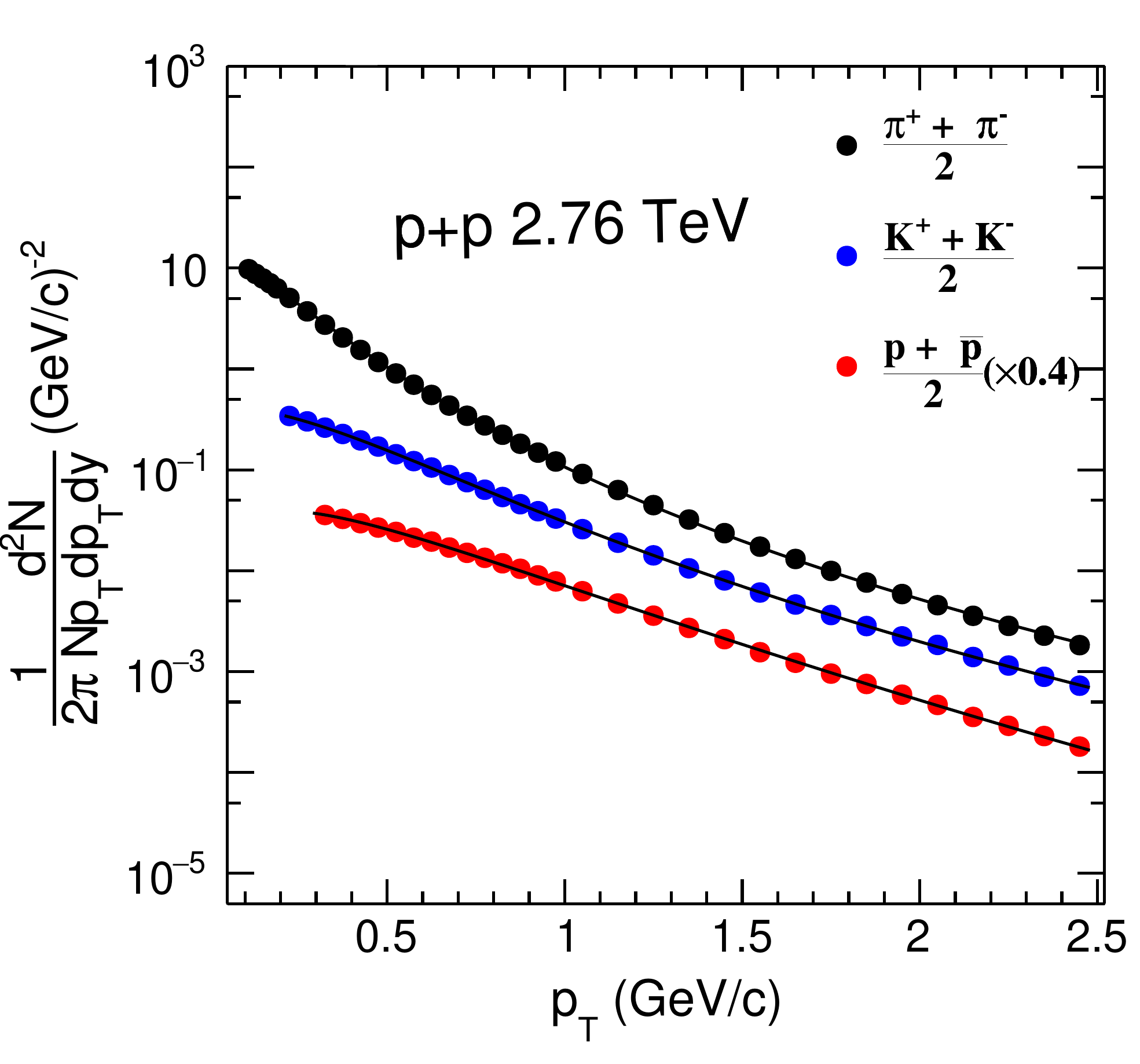}
\newline
\caption {(color online) Fitting of  invariant $p_{\rm T}$ spectra with Tsallis distribution using Eq.~\ref{eq6} for  
$\frac{\pi^ {+}+ \pi^{-}}{2}$ \cite{Abelev:2014laa}, $\frac{K ^ {+}+K ^{-}}{2}$ \cite{Abelev:2014laa}, $\frac{p+\bar p}{2}$ \cite{Abelev:2014laa} in $p+p$ collisions at $\sqrt{\mathrm{s}_{NN}}$=2.76 TeV. Symbols represent the  data points and line is the fit function.}
\label{fig4}
\end{center}
\eef

\bef[H]
\begin{center}
 \includegraphics[scale=0.28]{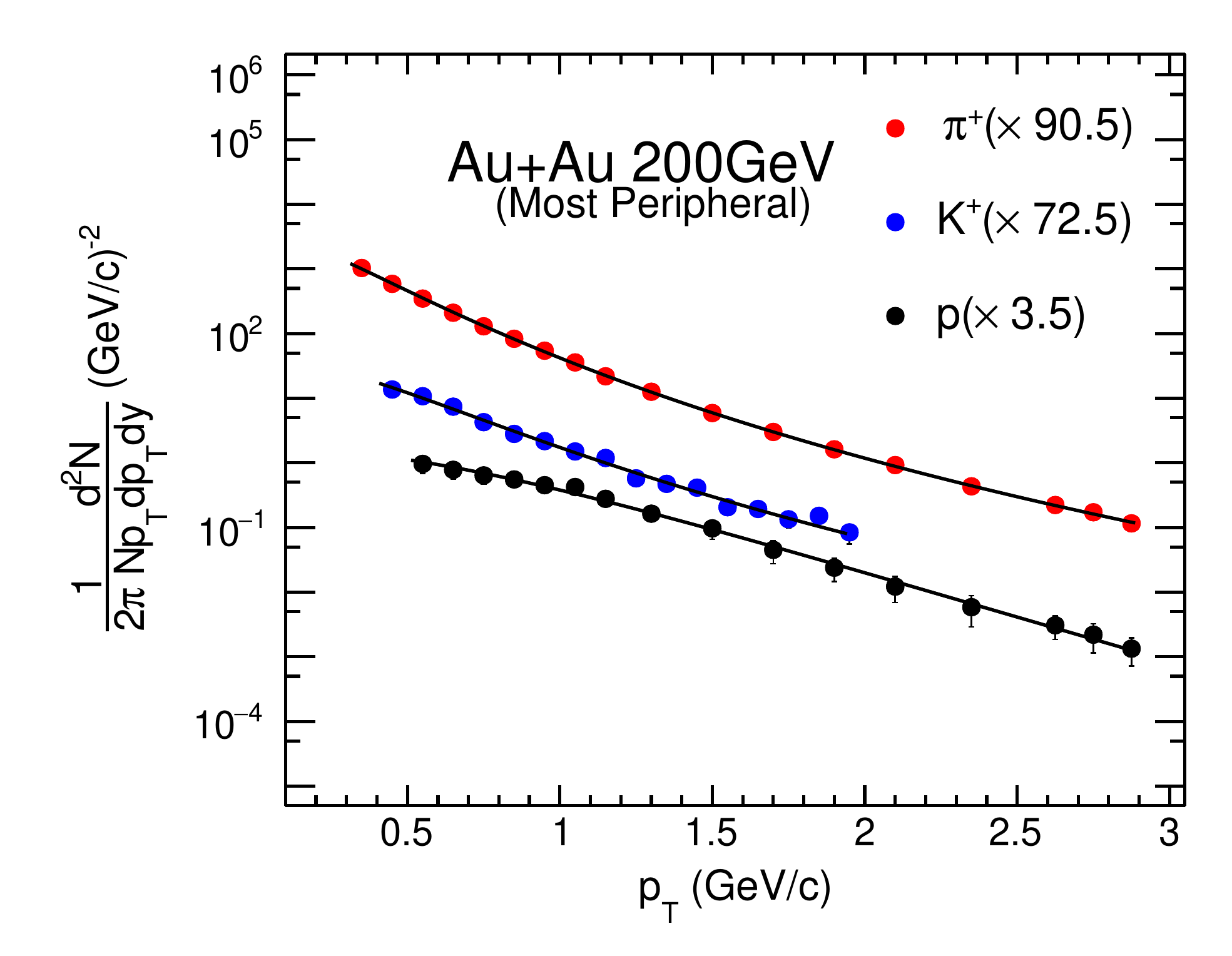}

\caption{(color online) Fitting of  invariant $p_{\rm T}$ spectra with Tsallis distribution including chemical potential using Eq.\ref{eq5} for $\pi^+$ \cite{Adler:2003cb}, $K^+$ \cite{Adler:2003cb}, $p$ \cite{Adler:2003cb} in most  peripheral Au+Au collisions at $\sqrt{\mathrm{s}_{NN}}$= 200 GeV. Symbols represent the  data points and the line is the fit function.} 
\label{fig9}
\end{center}
\eef

\bef[H]
\begin{center}
\includegraphics[scale=0.28]{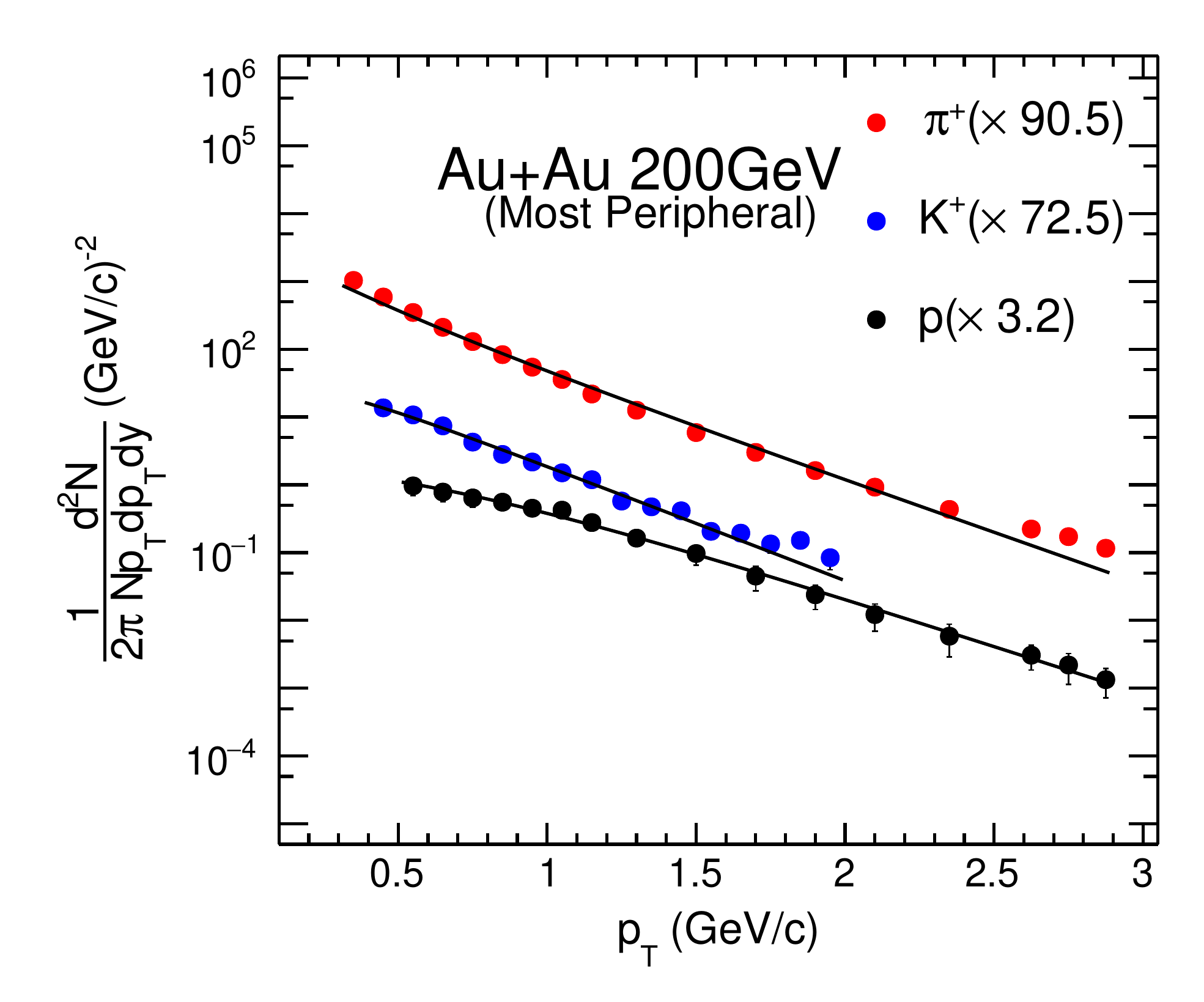}
\caption{(color online) Fitting of  invariant $p_{\rm T}$ spectra with Tsallis distribution including radial flow using Eq.~\ref{eq7} 
for  $\pi^+$ \cite{Adler:2003cb}, $K^+$ \cite{Adler:2003cb},
$p$ \cite{Adler:2003cb}  in most  peripheral Au+Au collisions at $\sqrt{\mathrm{s}_{NN}}$= 200 GeV.}
\label{fig14}
\end{center}
\eef

\bef[H]
\begin{center}
\includegraphics[scale=0.28]{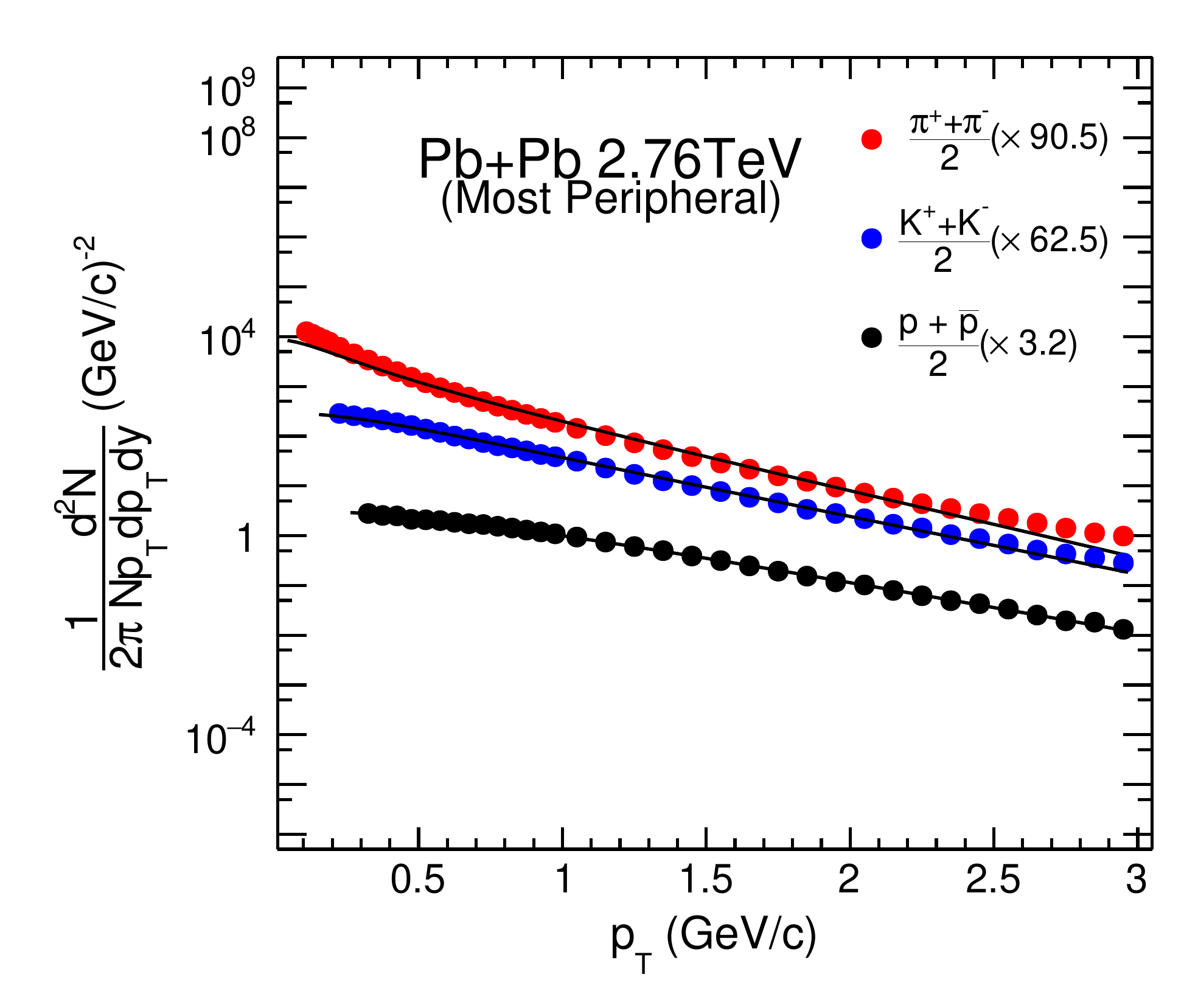}
\newline
\caption{(color online) Fitting of  invariant $p_{\rm T}$ spectra with Tsallis distribution including radial flow using Eq.~\ref{eq7} for $\frac{\pi^ {+}+ \pi^{-}}{2}$\cite{Abelev:2014laa}, $\frac{k ^ {+}+k ^{-}}{2}$\cite{Abelev:2014laa}, $\frac{p+\bar p}{2}$\cite{Abelev:2014laa} in most  peripheral Pb+Pb collisions at $\sqrt{\mathrm{s}_{NN}}$=2.76 TeV;
Symbols represent the data points and the line is the fit function. } 
\label{fig15}
\end{center}
\eef

\bef[H]
\begin{center}
\includegraphics[scale=0.28]{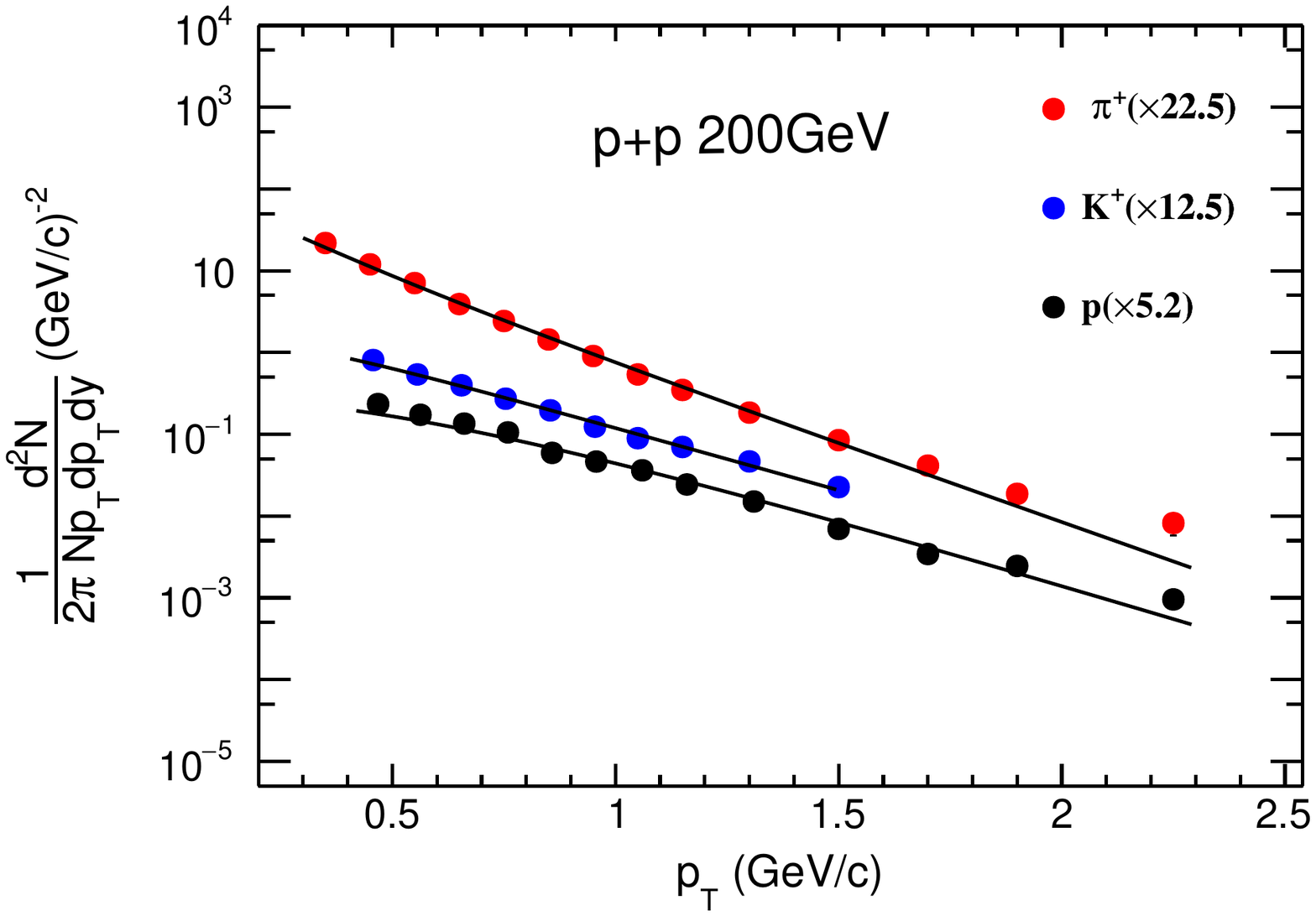}
\caption{(color online) Fitting of  invariant $p_{\rm T}$ spectra with Tsallis distribution including radial flow using Eq.~\ref{eq7} 
for $\pi^+$ \cite{Adams:2003qm}, $K^+$ \cite{Adams:2003qm}, $p$\cite{Adams:2003qm}  in $p+p$ collisions at $\sqrt{\mathrm{s}_{NN}}$= 200 GeV.  Symbols represent the data points and the line is the fit function. } 
\label{fig16}
\end{center}
\eef

\bef[H]
\begin{center}
\includegraphics[scale=0.24]{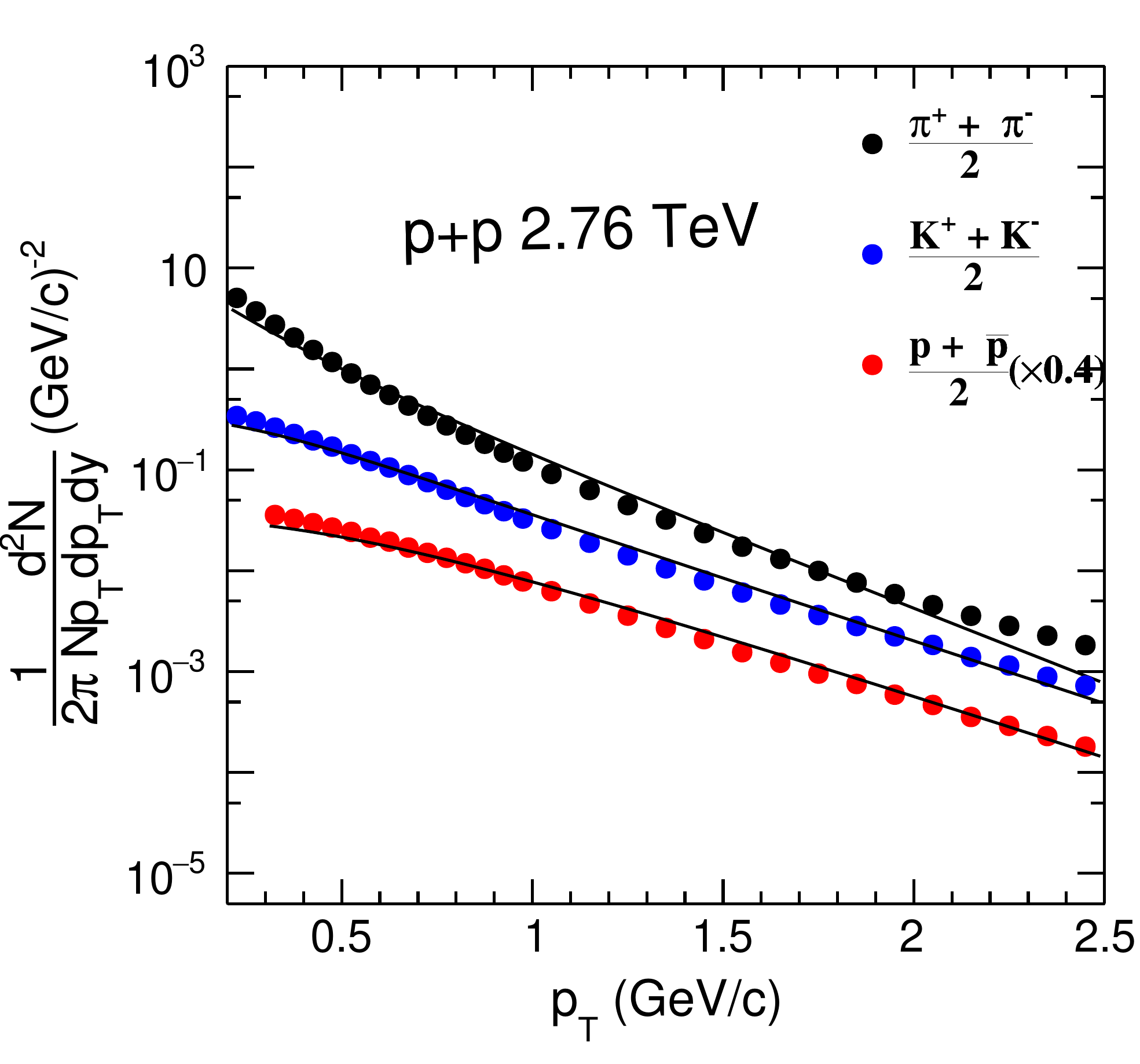}
\newline
\caption{(color online) Fitting of  invariant $p_{\rm T}$ spectra with Tsallis distribution including radial flow using Eq.~\ref{eq7} for $\frac{\pi^ {+}+ \pi^{-}}{2}$\cite{Abelev:2014laa}, $\frac{k ^ {+}+k ^{-}}{2}$\cite{Abelev:2014laa}, $\frac{p+\bar p}{2}$\cite{Abelev:2014laa} 
in $p+p$ collisions at $\sqrt{\mathrm{s}_{NN}}$=2.76 TeV.   Symbols represent the data points and the line is the fit function. } 
\label{fig17}
\end{center}
\eef


\end{document}